\pdfoutput=1
\documentclass{aa}  
\usepackage{graphicx}
\usepackage[colorlinks,urlcolor=blue,citecolor=blue,linkcolor=blue]{hyperref}

\usepackage{txfonts}

\usepackage{numprint}

\begin{document}

\title{EMISSA:\ Exploring millimetre indicators of solar-stellar activity}
\subtitle{ III. Comparison 
of Ca II indices and millimetre continua
in a 3D model atmosphere}
\titlerunning{Comparison of Ca II indices and mm continua}
\author{Sneha Pandit\inst{1,2}, Sven Wedemeyer\inst{1,2}, Mats Carlsson\inst{1,2}}
\institute{Rosseland Centre for Solar Physics, University of Oslo, Postboks 1029 Blindern, NO-0315 Oslo, Norway
   \and
    Institute of Theoretical Astrophysics, University of Oslo, Postboks 1029 Blindern, NO-0315 Oslo, Norway\\
\email{sneha.pandit@astro.uio.no}}

\abstract
  {
 Amongst several spectral lines, some of the strongest chromospheric diagnostics are offered by the Ca II H \& K lines. These lines can be used to gauge the temperature stratification of the atmosphere since the line core and wings are formed in different regions of the solar atmosphere. Furthermore, the Ca II lines act as tracers for the magnetic structure of the solar atmosphere, as the line cores are formed in the upper chromosphere even though they are formed in non-local thermodynamic equilibrium (NLTE). In contrast, the formation of millimetre (mm) continuum radiation occurs under  local thermodynamic equilibrium (LTE) conditions. As a result, the brightness temperatures obtained from observations with the Atacama Large Millimetre/Submillimetre Array (ALMA) offer a complementary perspective on the activity and thermal structure of stellar atmospheres. }  
  {
   The overall aim is to establish more robust solar/stellar activity indicators using ALMA observations in comparison with classical diagnostics, such as the s index and infrared triplet (IRT) index.
   }
  {
  We employed the 1.5D radiative transfer codes RH1.5D and advanced radiative transfer (ART)   to compute the synthetic spectra for the Ca II lines and the millimetre (mm) continua, respectively. These calculations were performed using an enhanced network atmosphere model, which incorporates non-equilibrium hydrogen ionisation generated by the state-of-the-art 3D radiation magnetohydrodynamics (rMHD) Bifrost code. To account for the limited spatial resolution of ALMA, we simulated the effect using a Gaussian point spread function (PSF). Additionally, we analysed the correlations and slopes of scatter plots between the Ca II indices and mm continuum for the original and degraded resolutions, focusing on the entire simulation box, quiet Sun regions, and enhanced network patches separately. The activity indices generated from these lines could further be used to compare the spectra of Sun-like stars with the solar spectrum.}
  {
  We present a comparative study between synthetic continuum brightness temperature maps at mm wavelengths (0.3 mm to 8.5 mm) and the Ca II activity indices; namely, the s index and infrared triplet (IRT) index.
  The Ca II activity indices and mm brightness temperatures are weakly correlated at the high resolution, with the highest correlation observed at a wavelength of 0.3~mm, corresponding to ALMA band~10. As the resolution decreases, the correlation consistently increases. Conversely, the slopes exhibit a decreasing trend with increasing wavelength, while the degradation of resolution does not noticeably affect the calculated slopes.}
   {As the spatial resolution decreases, the standard deviations of the Ca II activity indices and brightness temperatures decrease, while the correlations between them increase. However, the slopes do not exhibit significant changes. Consequently, these relationships could be valuable for calibrating the mm continuum maps obtained through ALMA observations.}
   
\maketitle

\section{Introduction}

For the Sun, as the closest star to Earth, spectra can be obtained at high spatial and temporal resolution. In contrast, the observed spectra of other stars are not as detailed but are the only basis for computing activity indices. These indices are generated by collapsing the resolved spectra by averaging, integrating, or convolving with a filter \citep{1989PASP..101..107S}. Activity indicators are calculated using spectral lines sensitive to the magnetic activity, as with the s index, a proxy based on the Ca II H and K flux integrated over the filters as defined by \citet{1978PASP...90..267V}, or the H$\alpha$ index as defined in \citet{2002AJ....123.3356G}. The abundance of spatiotemporally resolved data available for solar spectra allows for the use of these indices to retrieve information about the atmosphere. Furthermore, these indices can be used to compare the solar spectra with those of other stars to gain insights into their atmospheres and stratification.

The chromosphere is responsible for radiating away most of the energy that is transported to the outer solar atmosphere. The chromosphere is where the dynamical properties change from gas-pressure-dominated to magnetic force-dominated. The thermodynamical conditions in the chromosphere go from local thermodynamic equilibrium (LTE) to non-equilibrium conditions. The chromosphere is partially ionised, and the ionisation state changes from almost neutral to full ionisation \citep{2002ApJ...572..626C}. All these transitions make chromospheric physics very complex and it remains the most poorly understood region of the Sun \cite[see][ and references therein]{2016A&A...585A...4C}.

To study this layer and the temperature stratification of the Sun, Ca II H \& K lines are used as diagnostics. The line core for Ca II H \& K and the infrared triplet (IRT) are formed at chromospheric heights. The wavelengths away from the core are subsequently formed at lower layers and the wings are formed at deeper photospheric heights following local thermodynamic equilibrium (LTE). Observations in the Ca II H \& K lines imply a reversal of the atmospheric temperature profile with a chromospheric temperature rise. See Table~\ref{tab:formation heights} for the average formation heights as derived in the empirical study by \citet{2018A&A...611A..62B}. 

\begin{table*}
\centering
    
\begin{tabular}{|c|c|c|c|c|}
\hline

K line & Formation height range & Average formation height & H line & Average formation height below the K line \\
& (Mm) &(Mm)& & (km)\\
\hline
$K_3$ & 0.5-4.0 & 1.9 $\pm$ 0.6 & $H_3$ &150\\
$K_{2V}$ & 0.3-3.0 & 1.3 $\pm$ 0.5 & $H_{2V}$ &150\\
$K_{2R}$ & 0.3-3.0 & 1.0 $\pm$ 0.5 & $H_{2R}$ &100\\
\hline
\end{tabular}
    \caption{Formation heights of the features from the photosphere as described in \cite{2018A&A...611A..62B} based on Bifrost models derived from \citet{2016A&A...585A...4C}.}
    \label{tab:formation heights}
\end{table*}
 
The Ca II H and K lines are of interest due to their high depth and width resulting from the most probable transition between the highly populated 4s energy level and the immediate higher shell, 4p \citep{1990ASPC....9..103U, 2011A&A...528A...1W}. On the other hand, the infrared triplet of Ca~II  is between 3d and 4p.  It consists of three strong lines at 849.802, 854.209, and 866.214~nm, whose cores are formed in the chromosphere as well \citep{2018A&A...611A..62B, 1992A&A...254..258J}. They have smaller energy gaps and are brighter than the H~\&~K lines, as seen further in Fig.~\ref{fig:15k_vs_789grid}. The line core intensities for the IRT lines are higher than the H~\&~K line cores. The H and K lines are formed much higher in the less dense upper chromosphere where 3D and Partial Redistribution (PRD) effects play a more essential role in line formation than in the infrared triplet of Ca II \citep{1990prmc.book.....U}.

The Atacama Millimetre/submillimetre Array (ALMA) \citep{2009IEEEP..97.1463W} provides observations with a high spatial and temporal resolution for the millimetre (mm) continuum. The Ca II lines and mm continua both probe chromospheric heights. It is established that the continuum radiation observed at millimetre (mm) wavelengths is formed in the chromosphere \citep[see, e.g.,][, and references therein]{2016SSRv..200....1W}. This continuum radiation in the millimetre wavelengths (here 0.3–8.5\,mm) arises from free-free emission within the chromosphere, where the primary sources of opacity are H and H$^-$ free-free absorption \citep{1985ARA&A..23..169D}. These processes lead to an LTE source function that depends on the electron temperature of the plasma. Therefore, we can interpret the emergent intensity (the observed brightness temperature ($T_\mathrm{b}$) in the millimetre wavelength domain as local electron temperature \citep{2015A&A...575A..15L}. The opacity at mm wavelengths is directly dependent on the electron density, which deviates from equilibrium conditions in the chromosphere \citep[see, e.g.,][and references therein]{1985ARA&A..23..169D,2004A&A...419..747L,2016SSRv..200....1W}. While the formation process operates under LTE conditions, the non-equilibrium electron densities impact the diagnostic information derived from ALMA observations \citep[see, e.g.,][]{2019ApJ...881...99M, 2020A&A...643A..41D}. The millimetre continuum is nevertheless a useful diagnostic tool for probing the thermal properties of the atmospheres of the Sun and other stars as long as the non-equilibrium nature of the electron density is carefully taken into account.

Activity is a range of phenomena occurring in the outer layers of a star. Variations in the magnetic fields can be observed through variations of the observed intensity, which are used to calculate activity indices \citep{1975ApJ...200..747S,1999ApJ...515..812H}. The most commonly used indicator of chromospheric activity is the s index, the ratio of the flux in the core of the Ca II H and K lines to the nearby continuum \citep{1968ApJ...153..221W,1978PASP...90..267V}. The mm continua observations of the Sun can reveal the 3D structure of the gas and even the magnetic field in the outer layers as shown by \citet{2018A&A...620A.124D, 2016ApJ...825..138D}. The continuum wavelengths probe slightly different heights \citep{2015A&A...575A..15L}, giving us information of electron temperatures at different heights based on the mm wavelengths. This study aims to use ALMA's novel capabilities for a re-evaluation of the Calcium II activity indicators using a comparative solar-stellar study, with the solar model serving as a fundamental reference.

Solar observations from ALMA have been compared with the H$\alpha$ line by \citet{2019ApJ...881...99M, 2023FrASS...978405T, 2023A&A...673A.137P}, but there are fewer studies comparing mm solar data with Ca II lines. \citet{2020IAUS..354...42H} have compared nearly co-observed Ca~II~8542 data from SOLIS with ALMA data and reported striking similarities. \citet{2021ApJ...920..125M} compared co-observed IBIS Ca~II~8542 and H$\alpha$ data with ALMA Bands~3 and 6 data statistically to model the propagation of waves based on the velocities found with the Ca~II~8542 line. In \citet{2020A&A...643A..41D} the ALMA data was compared with IRIS Mg~II lines, and the authors recommended a closer comparison of ALMA data with NLTE lines like the Ca II lines. None of the studies mentioned have quantified the similarities between Ca~II and mm data for the Sun and this study is trying to fill that gap.

There has been a continuous effort undertaken to understand the activity based on the solar cycle. \citet{2011ApJ...736..114P} have found that the line core intensity and the line bisector at the line wings of Ca II 854.2~nm line are sensitive to the solar activity over 25 years covered in their study. \citet{2018csss.confE..37B} carried out a study to compare ALMA Band~6 data with theoretical models of different active phenomena. \citet{1993JGR....9812809D} studied the Mg~II index for the Sun over two solar cycles to measure long-term UV solar activity. The H Balmer lines variability based on the solar long-term variability has been addressed by \citet{2023ApJ...951..151C}. There are numerous studies that have employed the Sun-as-a-star approach. Our work takes this approach to understand the Ca~II indices in detail at different activity levels to add to the effort.

This work is focused on using and comparing Ca~II~intensities and millimetre brightness temperatures as complementary diagnostics. In this paper, we compare the mm continuum and Ca II activity indicators synthesised from a 3D realistic Bifrost model atmosphere to understand how these can be used as complementary diagnostics for the chromosphere. The model and the spectral synthesis are described in Sect.~\ref{Sec:Methods}. In Section~\ref{Sec:Results}, we present the quantitative comparison of the two diagnostics which is further qualitatively discussed in Sect.~\ref{Sec:Discussion}. The study is concluded in Sect.~\ref{Sec:Conclusions}.

\section{Methods}
\label{Sec:Methods}

A snapshot of a numerical 3D simulation of the solar atmosphere (see Sect.~\ref{Sec:Bifrost}) was used as input for radiative transfer codes for the synthesis of Ca II lines (see Sect.~\ref{Sec:RH}) and the continuum intensity at millimetre wavelengths (see Sect.~\ref{sec:ART}). The resulting maps for different wavelengths were then degraded to a spatial resolution typically achieved with ALMA, as described in Sect.~\ref{sec:psf} before the activity indicators were determined (see Sect.~\ref{sec:activity_indices}) and compared to the mm brightness temperatures. 

\subsection{Bifrost}
\label{Sec:Bifrost} 

Bifrost is a 3D simulation code that provides an unprecedented view of the solar atmosphere with high resolution \citep{2011A&A...531A.154G}. The computational box measures 24 Mm x 24 Mm horizontally and extends 2.4 Mm below the visible surface and 14.4 Mm above, covering the upper regions of the convection zone, photosphere, chromosphere, transition region, and corona \citep{2016A&A...585A...4C}. It consists of 504 x 504 x 496 grid points with a horizontal resolution of 48 km and a variable grid separation in the vertical direction ranging from 19 km in the photosphere and chromosphere up to 5 Mm height and then increasing to 100 km at the top boundary. Both top and bottom boundaries are open, and the magnetic field is passively advected at the bottom boundary, with no additional field fed into the computational domain \citep{2011A&A...531A.154G, 2016A&A...585A...4C}. The name of the model used in this project is en024048, with 'en' standing for enhanced network, the horizontal size of the computational domain 24~Mm, and the horizontal grid size 48~km (more relevant details are discussed in \citet{2023A&A...673A.137P} and references therein). 

As seen in \cite{2020A&A...635A..71W}, the simulation has different regions. At the lateral boundaries, it is more quiet and hence, less magnetically active. The en024048 simulation features an enhanced network (EN) patch with loops in the middle of the computational domain, whereas the outer parts are more representative of magnetically less active, quiet Sun (QS) conditions \citep[see Fig.~1 in ][for more details about the temperature and absolute magnetic field]{2023A&A...673A.137P}. This diverse behavior exhibited by the model atmosphere is reflected in the varying intensity of the Ca~II lines (as seen in Fig.~\ref{fig:Ca_II_maps}).

\subsection{Radiative Transfer (RH) code}
\label{Sec:RH}

The Ca~II lines are synthesised from the Bifrost model atmosphere using the radiative transfer code RH1.5D. The multilevel atoms and molecules are solved for statistical equilibrium and radiative transfer equations in a self-consistent manner using the RH code, as detailed in \cite{2015A&A...574A...3P, 2001ApJ...557..389U}. The formation of the Ca~II~H and K wings occurs in the photosphere under LTE, while the cores form in the chromosphere, where the low densities cause significant departures from LTE. If the LTE approximation is used to synthesise line cores, it leads to unrealistic emission profiles since the temperature increases in the layers above the temperature minimum. \cite{1990ASPC....9..103U} has emphasised that PRD effects are crucial for the formation of these lines. Thus, the RH's non-LTE PRD capability is employed to synthesise spectra for the Ca II H \& K lines, 
similarly to what is seen in the studies from \cite{2019A&A...631A..80B, 2017A&A...599A.118B}.

\begin{figure*}[ht!]
\centering
\vspace*{-30mm}
\includegraphics[width=1.0\textwidth]{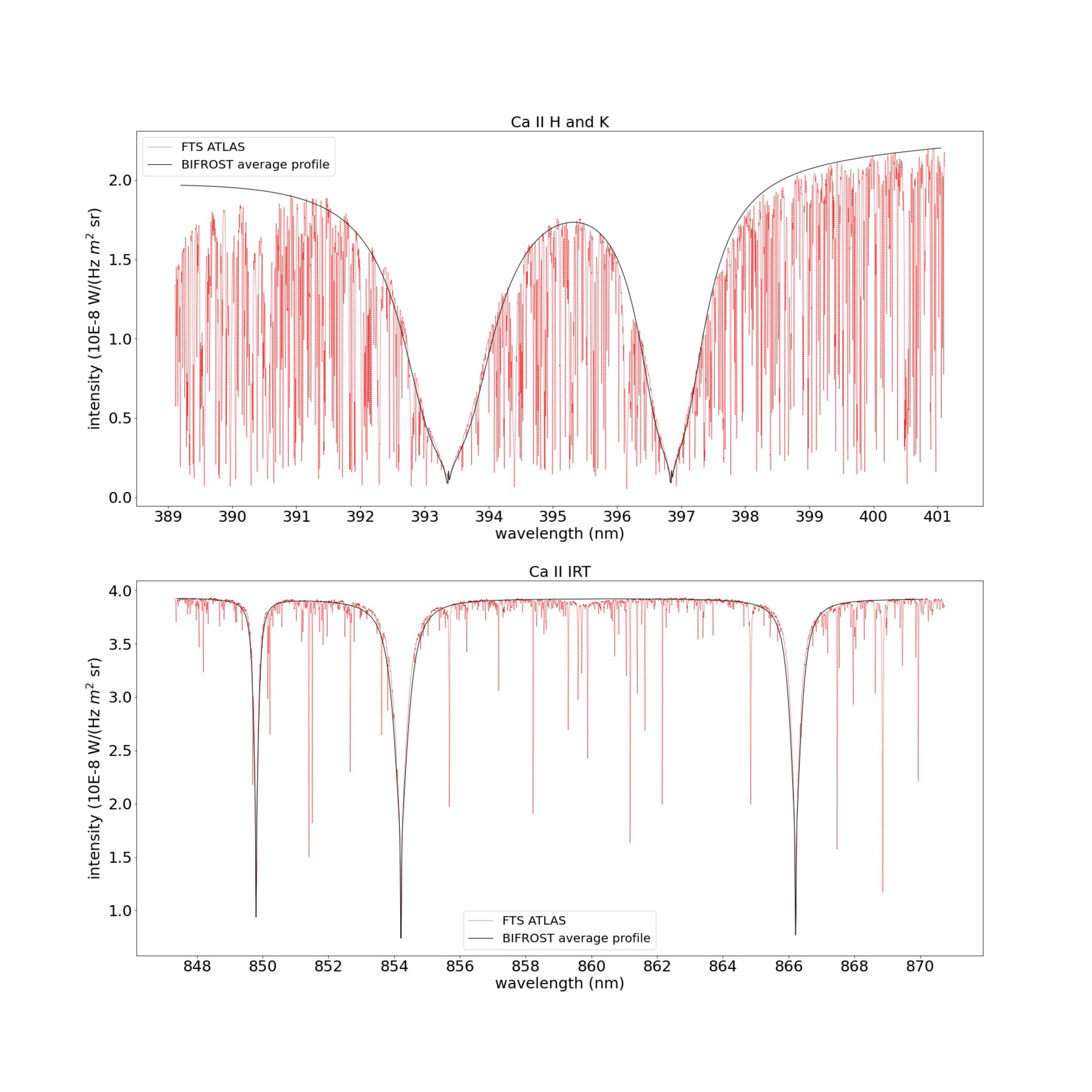}
\vspace*{-20mm}
\caption{FTS solar ATLAS data in both Ca II H and K band and IRT band is shown (in red) in comparison with the average Ca II data (in black) generated from the Bifrost model without line blanketing included.}
\label{fig:2bands_comparison_ATLAS_CaII_run}
\end{figure*}

\begin{figure*}[ht!]
\centering
\includegraphics[width=0.8\textwidth]{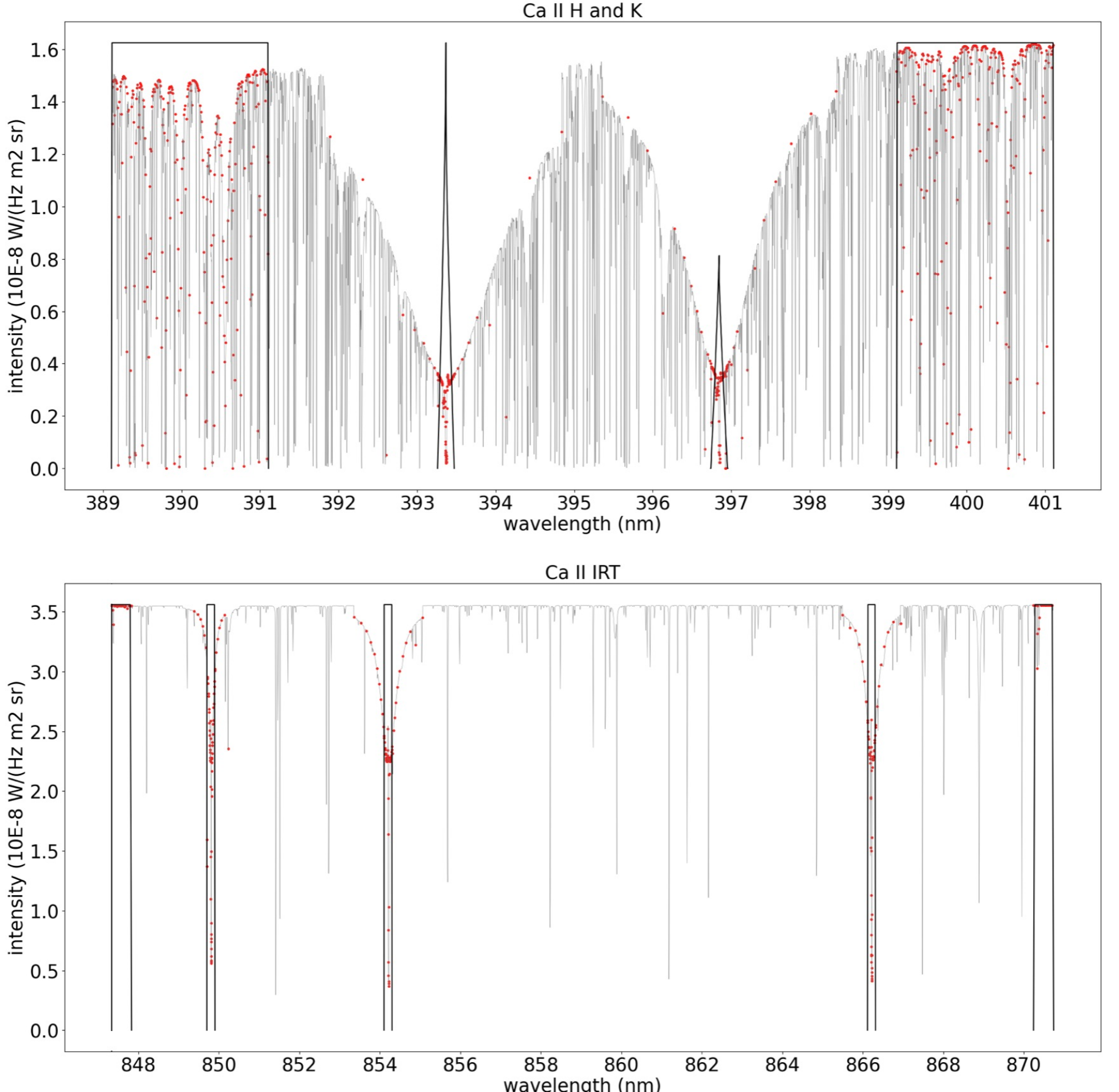}
\caption{Resolved spectra simulated with all Kurucz lines in the range 389.1067~nm to 401.1067~nm and 847.33~nm to 870.74~nm shown in black fine lines in comparison with the run considered in the study with comparatively fewer Kurucz lines shown in red dots for column (0,0), the filters for the different bands required for calculating the indices are shown in thick black lines.}
\label{fig:15k_vs_789grid_2bands}
\end{figure*}

\begin{figure*}[ht!]
\centering
\vspace*{-20mm}
\includegraphics[width=1.0\textwidth]{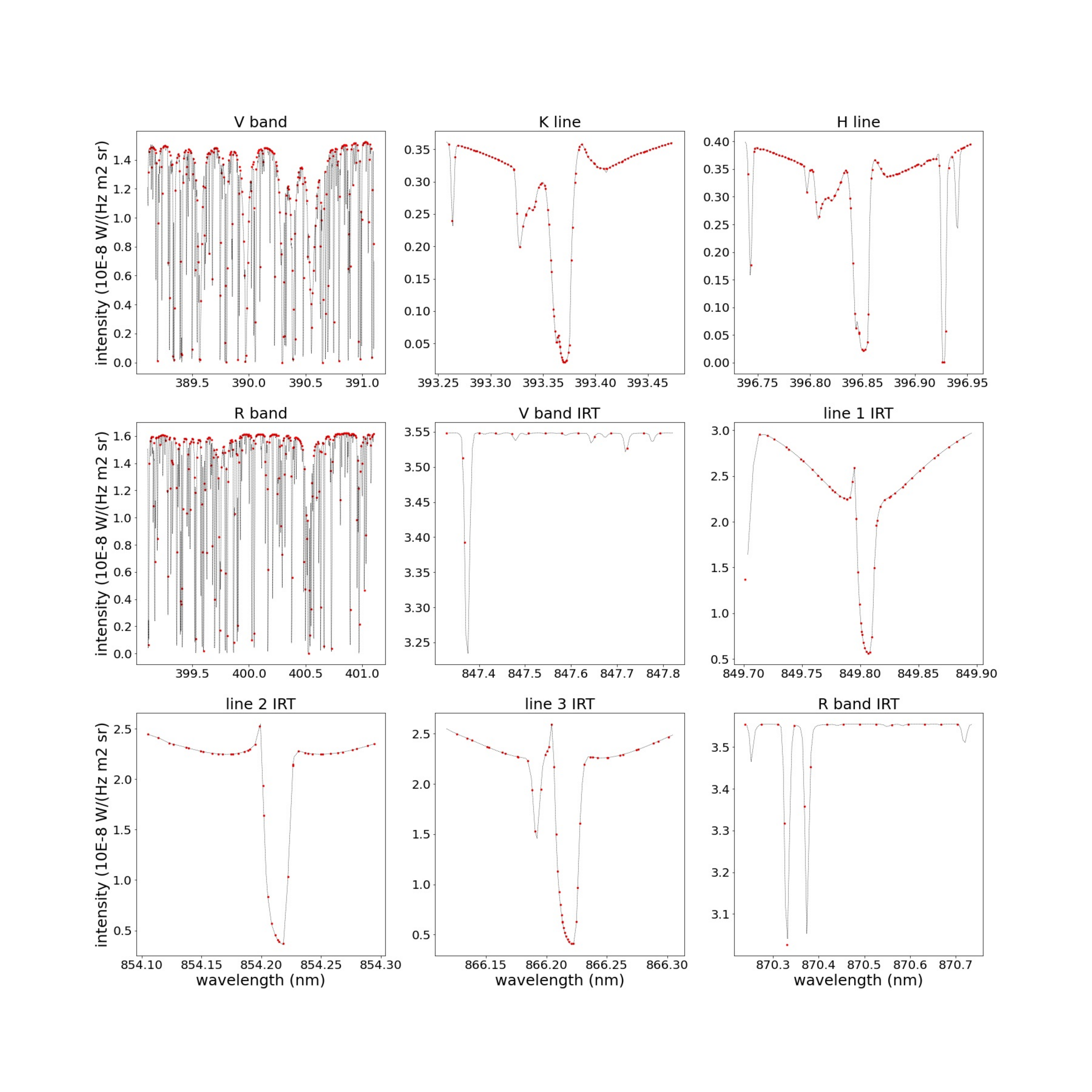}
\vspace*{-20mm}
\caption{Same legend as in Fig. \ref{fig:15k_vs_789grid_2bands}, for the bands of interest for the calculation of the s index and IRT index. The panels show these bands: V band (389.1067~nm to 391.1067~nm), K line (393.3664~nm), H line (396.847~nm), R band (399.1067~nm to 401.1067~nm), V band for IRT (847.33~nm to 847.83~nm), 1st triplet line (849.8~nm), 2nd triplet line (854.2~nm), 3rd triplet line (8662~nm), and R band for IRT (870.24 nm to 870.74~nm).}
\label{fig:15k_vs_789grid}
\end{figure*}

Figure~\ref{fig:2bands_comparison_ATLAS_CaII_run} shows the agreement between the averaged Ca~II data synthesised using snapshot 420 of the same Bifrost model 'en024048', which is --- minutes more evolved than the snapshot used in the study. This data is used to save computational time as it was produced for another work. 
As can be seen in the figure, the average Ca~II line profiles follow the ATLAS profiles quite neatly, especially in the (photospheric) line wings. The average Bifrost line profiles were produced without including line blanketing to speed up the calculations when calculating the spectra for all 504x504 columns. When calculating the indices in the paper, we do include line blanketing in a sparse grid (as explained in Figs. \ref{fig:15k_vs_789grid_2bands} and \ref{fig:15k_vs_789grid}).

The Kurucz line list\footnote{\url{https://lweb.cfa.harvard.edu/amp/ampdata/kurucz23/sekur.html}} has an extensive list of background lines which contribute to the opacity. However, lines with lower transition probability have an insignificant impact on the intensity.  To reduce the required computation time, which is substantial for a model with 504x504 columns,  the lines affecting the calculated s index the most were selected for the test run.  The criterion for selection for this list was that the resulting s index is within 1\% of the calculated s index with the whole line list for each 50th column (121 columns in total).


In Figure~\ref{fig:15k_vs_789grid_2bands}, the intensity is calculated with all the background lines in the Kurucz line list. The calculation of the s index requires the lines within the predefined triangular filters and continuum bands from \cite{1978PASP...90..267V} on either side (red and blue) of the lines shown in the figure by solid black lines. The same is also shown for the IRT index. An alternative method of saving computational power is to alter the computational wavelength grid and calculate only the required intensities within the filters. The red dots show the spectra generated with the finalised grid for the final run. This wavelength grid table was also chosen for the final run as it was observed that the calculated s index is within 1\% of the calculated s index with the whole grid.

The outcome of both the runs for the wavelength ranges used for calculating the s index and IRT index in the first column (0,0 column) is shown in Fig.~\ref{fig:15k_vs_789grid}. The panels show the following bands: V band, K, H, R bands, V band for the infrared triplet, first triplet line (849.8~nm), second triplet line (854.2~nm), third triplet line (866.2~nm), and R band for IRT. The grey lines show the runs with the whole Kurucz line list and correspondingly saturated wavelength grid. The almost constant continuum intensity is due to the lack of strong lines so this constant value at 35.6 nW/(Hz m2 sr) is nearly the true continuum. Red dots show the run with the finalised grid and the Kurucz line lists truncated at the 99th percentile. The lines are in good agreement, and the blue and red bands that indicate the denominators of the respective indices are reaching the same heights in intensity at the continuum. The averaged synthetic line profiles generated from the Bifrost model are observed to follow the FTS ALTAS spectra.

\begin{figure*}[ht!]
\centering
\vspace*{-15mm}
\includegraphics[width=1.0\textwidth]{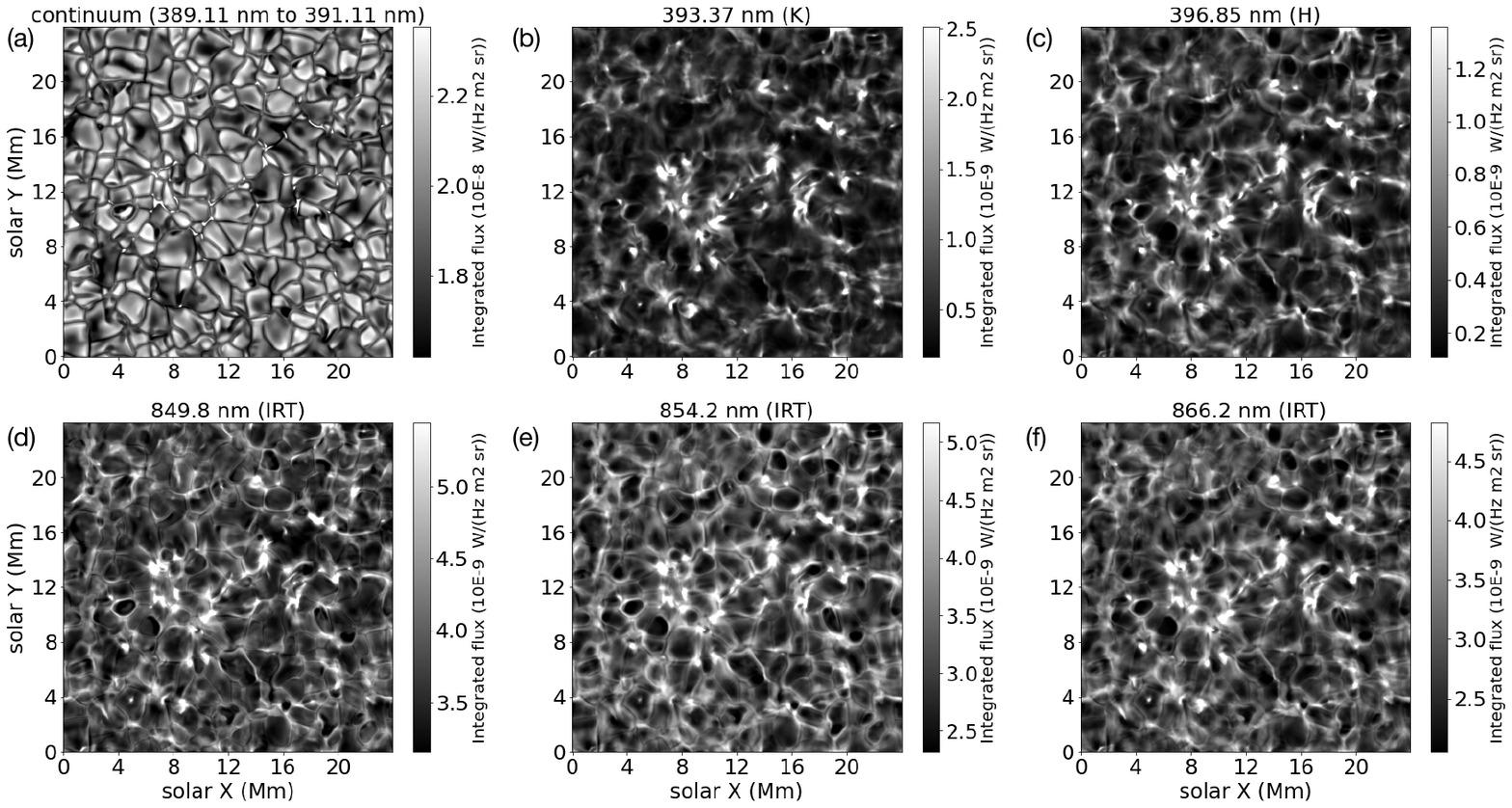}
\vspace*{-25mm}
\caption{Maps showing the integrated intensities, within the 1st and 99th percentile, for the wavelength ranges of interest for the calculation of the s index and the IRT index, namely, the continuum, triangular bandpasses of $2\AA$ centred at lines K and H, and rectangular bandpasses of $2\AA$ around lines in the infrared triplet.}
\label{fig:Ca_II_maps}
\end{figure*}

Figure~\ref{fig:Ca_II_maps} shows the intensity maps for different wavelengths formed in different layers from the Bifrost simulation box. The continuum is formed in the photosphere which is seen from the synthetic observations. The H and K line cores on the other hand show the footpoints of the enhanced network which is in the chromosphere. Similarly in the infrared triplet lines, the inter-network and network regions are seen prominently.

\subsection{Ca~II~IRT observations of $\alpha$ Cen A}
\label{sec:Alf_cen_A_data}
In this study,  \ion{Ca}{II}~IRT observations of the solar-like star $\alpha$~Cen~A are used for comparison. The data is acquired using MUSICOS (\textit{MUlti SIte COntinuous Spectroscopy}) high-resolution echelle spectrograph (R=35000, $\delta\lambda$ 0.024~nm at $\lambda$ 850~nm) mounted at the 1.6~m telescope of the Observatório do Pico dos Dias (OPD), operated by the Laboratório Nacional de Astrofísica (LNA/MCTI), Brazil from \citet{2016A&A...595A..11L}. The wavelength range covers approximately 550-880\,nm divided into 50 spectral orders. The $\alpha$~Cen~A \textit{echelle} spectrum is reduced following the standard  Image Reduction and Analysis Facility (IRAF) procedures: bias subtraction, flat field, scattered light corrections, 1D extraction, and Doppler and blaze function corrections. The continuum in the processed spectra around the Ca IRT lines (834.0~nm and 877.1~nm) is normalised to unity using low-order Legendre polynomials by \citet{2016A&A...595A..11L}.

\subsection{Synthesis of millimetre continuum brightness temperatures} 
\label{sec:ART}

Using the advanced radiative transfer (ART) code developed by \citet{2021_art}, the intensity of the continuum radiation at millimetre wavelengths is computed for the 3D input model described in Sect.~\ref{Sec:Bifrost}. The output intensities ($I_{\lambda}$) are then converted into brightness temperatures ($T_b$) by applying the Rayleigh-Jeans approximation, as shown in the following equation:
\begin{equation}
T_b=\frac{\lambda^4}{2k_Bc}I_{\lambda},
\end{equation}
Here, $\lambda$, $k_B$, and $c$ represent the wavelength, Boltzmann constant, and speed of light, respectively. This process is repeated for 34 wavelengths, covering a range that can be observed using the available and potential receiver bands of ALMA, as demonstrated in \citet{2023A&A...673A.137P}.

\subsection{Image degradation}
\label{sec:psf}

We adopted the same procedure for image degradation described in Section 2.6 of \citet{2023A&A...673A.137P} to degrade the Ca II data before computing the integrated intensities and indices. The synthetic beam's exact shape and width can vary even for observations within the same receiver band depending on the array configuration, and thus the diameter of the synthesised aperture and the position of the source in the sky with respect to the array baseline. For instance, the band 3 data sets available in the Solar ALMA Archive \citep[SALSA,][]{2022A&A...659A..31H} have widths of approximately 2", including the beam with a width of 1.92'' $\times$ 2.30'' for the 2016.1.01129.S data set. To simplify this study, the beam's width (i.e., the point spread function, PSF) is set to 2" at a wavelength of $\lambda = 3.0$ mm and then linearly scaled as a function of the observing wavelength $\lambda$. The wavelength range we considered includes the receiver bands that are currently available for solar observations with ALMA (bands 3, 5, and 6, i.e., 2.59--3.57~mm, 1.42--1.90~mm, and 1.09--1.42~mm, respectively), as well as bands 1, 2, 4, and 7-10, covering the entire range from 0.3 mm to 8.5 mm that may be available in the future. This assumes that the interferometric array is a large circular dish with a full width at half maximum (FWHM) in the x-y directions of $2" \times (\text{wavelength in mm})/3\text{ mm}$ for the synthetic Gaussian kernel. A circular wavelength-dependent Gaussian point spread function (PSF) is then applied to both the millimetre brightness temperature maps and the Ca II intensity maps.

\subsection{Semi-empirical models}

The 1D semi-empirical reference models representative of quiet Sun conditions VAL~C \citep{1981ApJS...45..635V} and FAL~C \citep{1993ApJ...406..319F}, the plage model by \citet{2009ApJ...707..482F}, and the sunspot model by  \citet{1994ASIC..433..169S} as provided by Loukitcheva (priv. comm.) are used to compare with the 3D results. Corresponding Ca II line profiles and mm continuum intensities for these 1D semi-empirical atmosphere models are calculated with the radiative code RH~1.5D (as described in Sect.~\ref{Sec:RH}).

\subsection{Calculation of activity indices}
\label{sec:activity_indices}

From the intensities calculated in the run set for the whole simulation box, namely, 504 $\times$ 504 columns (see Sect.~\ref{Sec:RH}), the calcium activity indices were calculated. The definition from \cite{1978PASP...90..267V} was used. Figure~\ref{fig:Ca_II_maps} shows the individual integrated intensities over the defined filters as per \citet{1978PASP...90..267V} for the components of the calcium s index. The continuum intensity image shows photospheric granulation, where the V band has slightly lower continuum intensity than the R band as expected from the wavelength-dependence of the solar flux spectrum. The H and K line-integrated intensities over the triangular filter centred at the line core show the internetwork regions and the enhanced network loop footpoints with significantly higher intensities. By definition of the s index, the triangular filter for the H line has half the height of that of the K line, namely, equal to the continuum. The s index is calculated by using the standard definition from \cite{1978PASP...90..267V} as: \begin{equation}
    S= {\alpha_c} \times 8 \times \frac{F_H+F_K}{F_V+F_R},
\label{eq:s_index}\end{equation} 
where $F_H$ is the integrated intensity over a triangular filter between 396.738\,nm and 396.956\,nm, centred at 396.847~nm and $F_K$ is the integrated intensity over a triangular filter between 393.2574~nm and 393.4754~nm, centred at 393.3664~nm, and $F_V$ (389.1067~nm to 391.1067~nm) and $F_R$ (399.1067~nm to 401.1067~nm) are the averaged intensities integrated over the rectangular filter over the whole box \citep{2021ApJ...914...21S}. The factor of 8 in Eq.~(\ref{eq:s_index}) was introduced to account for the differences in integration time in the different filters centred at the line cores as compared to the continuum observations as described in \citet{2007AJ....133..862H}. The factor is kept in the definition to allow for comparison with other published values. The instrumental calibration factor $\alpha_c$ is set to a value of 2.4 for the Mount Wilson Observatory \citep{1978PASP...90..267V} but it is set to 1 in this study as the synthetic data is not subject to instrumental properties.

The Ca infrared triplet index is defined based on the definition of the s index \citep{2014MNRAS.444.3517M}:\begin{equation}
     \text{CaIRT index}=\frac{F_{8498} + F_{8542} + F_{8662}}{F_{V_{IRT}} + F_{R_{IRT}}},
\label{eq:IRT}
\end{equation} 
where $F_{8498}$, $F_{8542}$ and $F_{8662}$ are the fluxes measured in 2~\AA~ rectangular bandpasses centred on the respective Ca II IRT lines, while $F_{V_{IRT}}$ and $F_{R_{IRT}}$ are the fluxes in 5~\AA~ rectangular bandpasses centred on two continuum points, 8475.8~\AA~ and 8704.9~\AA, at either side of the IRT lines \citep[][Eq. 1]{2013LNP...857..231P}.

\section{Results}
\label{Sec:Results}

\subsection{The s index}
\label{sect:s_index}

\begin{figure*}[ht!]
\centering
\vspace*{-15mm}
\includegraphics[width=1.0\textwidth]{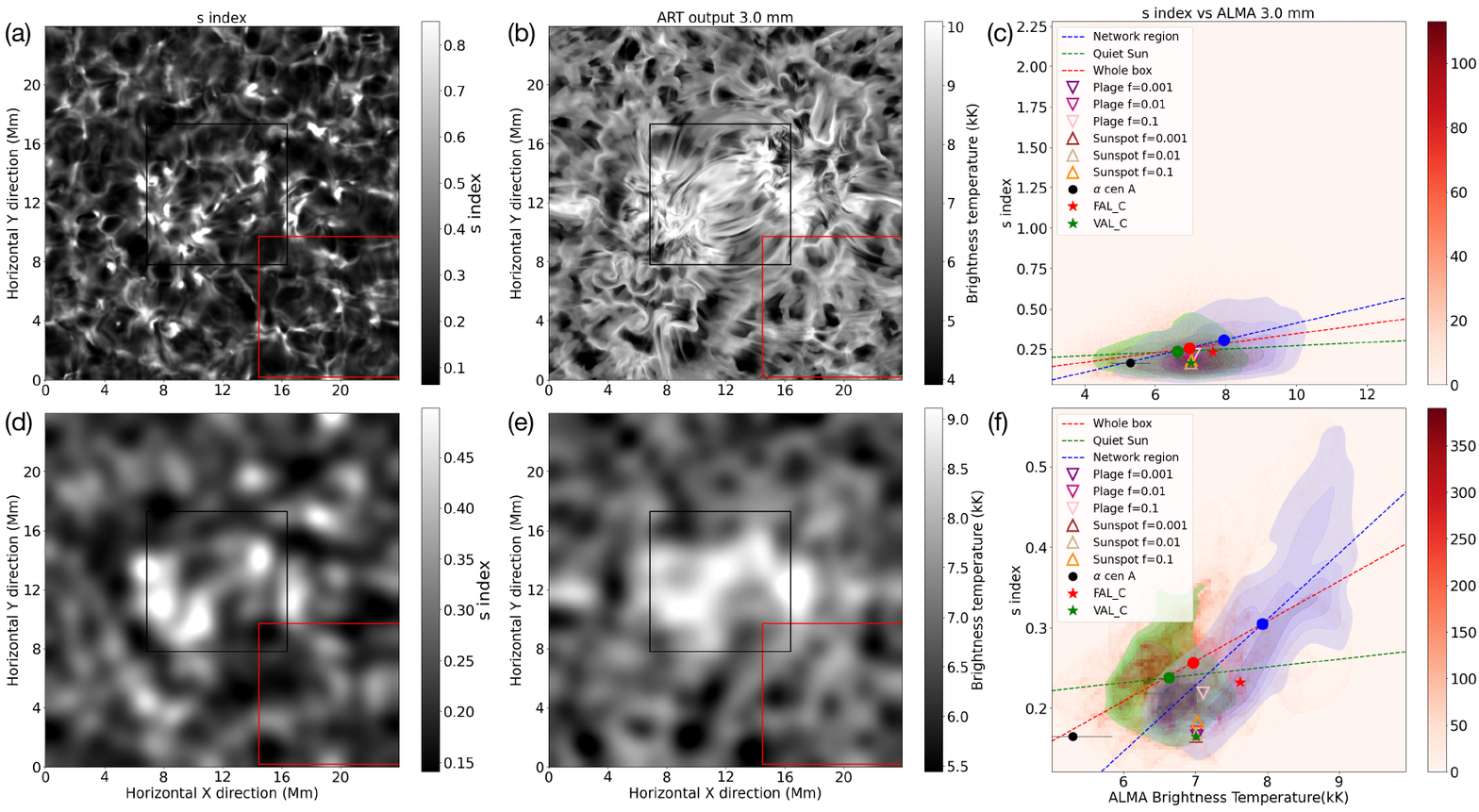}
\vspace*{-20mm}
\caption{Comparisons of the s index (a,d) and ALMA at 3.0\,mm brightness temperature maps (b,e) at original (top row) and degraded (bottom row) resolution. The black and red boxes on these maps denote the chosen EN and QS regions respectively. The last column is a correlation contour plot between the s index and ALMA brightness temperature at 3.0\,mm, with contours with linear fits (dashed lines) and means (circles) of the three cases: the whole simulation box in red, the QS box in green and the EN region box in blue. The red and green star data points are for the FAL~C and VAL~C 1D semi-empirical models and the black circle with error bars is an observational data point for G2V type star $\alpha$ Cen A. The 1D solar semiempirical data with respective filling factors using \citet{1994ASIC..433..169S} sunspot model and \citet{2009ApJ...707..482F} Plage model H along with VAL C data are represented by upwards facing triangles and downwards facing triangles.}
\label{fig:CaII_ALMA_3.0mm_comparison_s_index}
\end{figure*}

As seen in Fig.~\ref{fig:Ca_II_maps}a, the photospheric granulation is visible in the integrated intensities in the continuum. In panels b and c, the integrated intensities for K and H lines using the triangular filters defined by \citet{1978PASP...90..267V} are shown respectively. The s index calculated from these integrated intensities is shown in Fig.~\ref{fig:CaII_ALMA_3.0mm_comparison_s_index}a. Figure~\ref{fig:CaII_ALMA_3.0mm_comparison_s_index} shows the comparison between the EN region and QS region for the original and the degraded resolution corresponding to the assumed ALMA resolution at a wavelength of 3~mm. Panel~a shows the s index calculated as per Eq.~\ref{eq:s_index} based on the procedure described in Sect.~\ref{sec:activity_indices}. The shock pattern from the lower chromosphere is observed from this map similar to the K and H lines in Fig.~\ref{fig:Ca_II_maps}~b,c. 
On the other hand, the 3~mm map (panel~b)\ shows a layer at heights higher than the height represented by Ca II H\&K integrated intensity shown in Fig.~\ref{fig:Ca_II_maps}~b,c. The maps appear more similar at reduced resolution after convolution with a circular Gaussian kernel corresponding to the spatial resolution at a wavelength of 3~mm, namely, with a PSF of 2" ~(see panels d and e). The black and red boxes on these maps denote the chosen EN and QS regions for further comparison. The EN region is chosen to be centred at the enhanced network as seen in the degraded resolution, and the QS region is placed at the bottom right corner of the simulation box. Both boxes are of the same size of 9.5~Mm $\times$ 9.5~Mm (200~pixels $\times$ 200~pixels) \citep[similar to the analysis in][]{2023A&A...673A.137P}.

Following the same procedure as described in \citet{2023A&A...673A.137P}, Fig.~\ref{fig:CaII_ALMA_3.0mm_comparison_s_index} presents scatter contour plots for three different sets of pixels: the entire simulation box, the QS box (indicated by a red square in panels a, b, d, e), and the EN box (indicated by a black square in panels a, b, d, and e). The distribution for the entire simulation box is depicted by red contours, accompanied by a colour bar indicating the number of pixels within each bin. The QS and EN distributions are represented by green and blue contours, respectively. The distributions at the original resolution are substantially more spread out than the distributions at degraded resolution because of the associated loss of variations on small spatial scales. This effect is evident from the visual comparison of the s index maps in panels a and d, as well as the mm brightness temperature maps in panels b and e. The average values for the whole box, QS, and EN are denoted by red, green, and blue points, respectively, in panels c and f. The mean s index values for the whole box, QS and EN regions, 0.257, 0.237, and 0.307 are shown by a circle corresponding to the colours mentioned before. The s index value depends on the activity level as there are notable differences in these means.

Furthermore, for comparison, the s index and the $T_b$ at the mm wavelengths are plotted for the VAL~C and FAL~C semi-empirical model atmospheres. The VAL~C model has consistently lower brightness temperature than the FAL~C model as seen in Fig.~6 in \citet{2023A&A...673A.137P}. Also, both these semiempirical models exhibit lower s indices than the calculated averages for the whole box, EN and in the case of the VAL~C model, even lower than QS. This issue is further addressed in Sect.~\ref{Sec:Discussion}. When plotted on the s index vs temperature planes, these quiet Sun models lie in between the EN and QS distributions, demonstrating the agreement between the data from our study and the semi-empirical models similar to \citep[see Sect.3.1 in][]{2023A&A...673A.137P}. 

\begin{figure*}[ht!]
\centering
\vspace*{-15mm}
\includegraphics[width=1.0\textwidth]{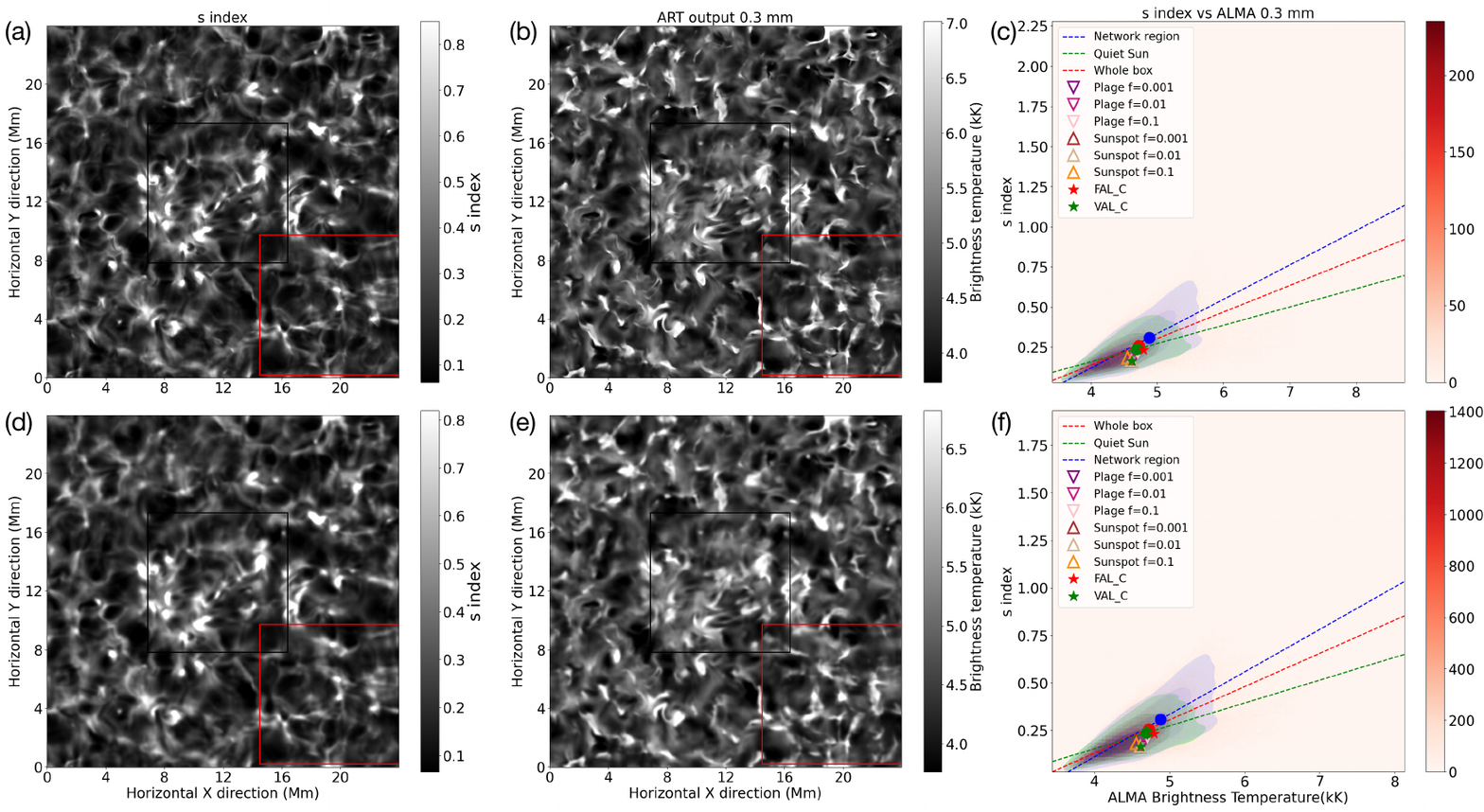}
\vspace*{-20mm}
\caption{Same as Fig.~\ref{fig:CaII_ALMA_3.0mm_comparison_s_index} but for a wavelength of 0.3~mm.}
\label{fig:CaII_ALMA_0.3mm_comparison_s_index}
\end{figure*}

\begin{figure*}[ht!]
\centering
\vspace*{-15mm}
\includegraphics[width=1.0\textwidth]{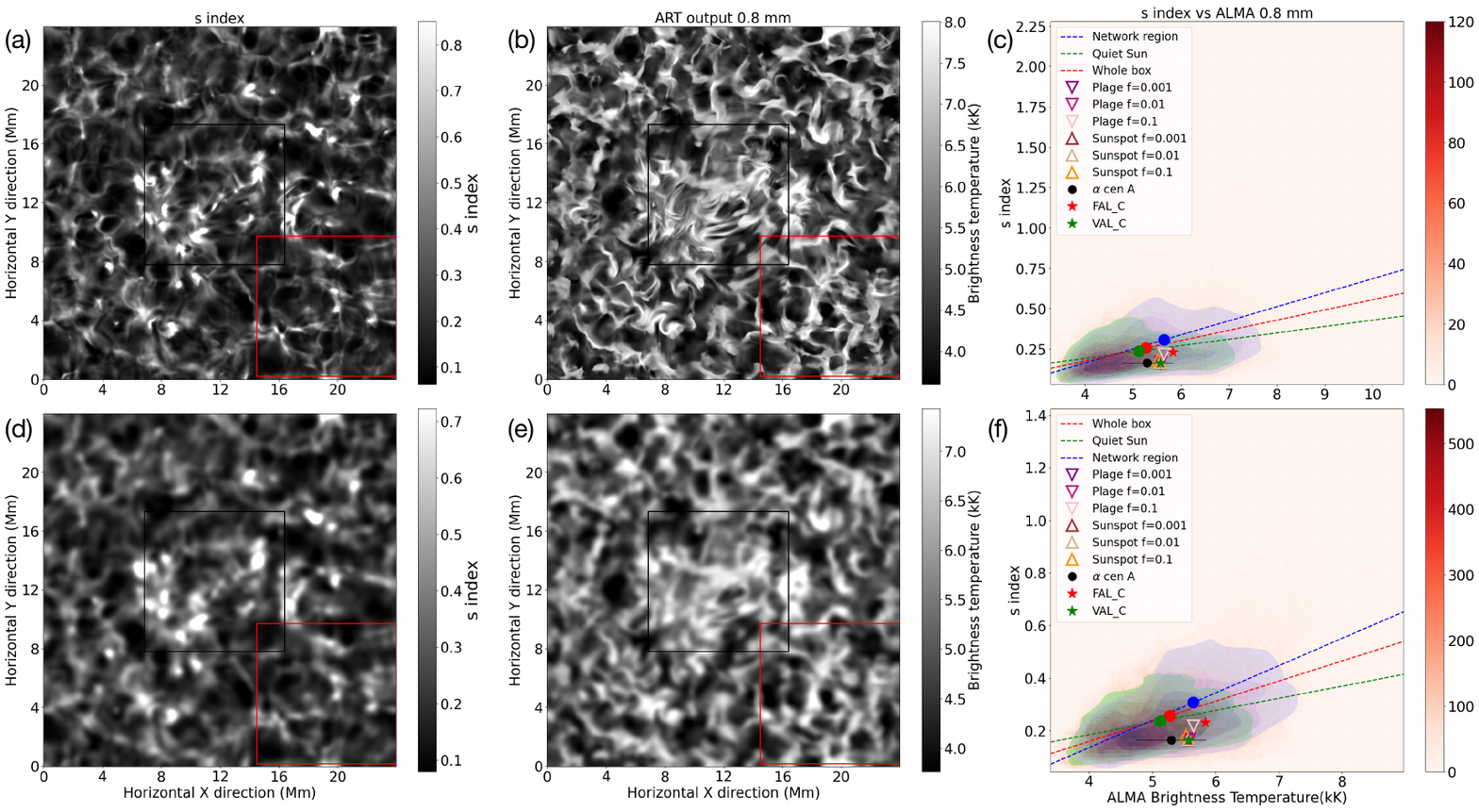}
\vspace*{-20mm}
\caption{Same as Fig.~\ref{fig:CaII_ALMA_3.0mm_comparison_s_index} but for a wavelength of 0.8~mm.}
\label{fig:CaII_ALMA_0.8mm_comparison_s_index}
\end{figure*}

\begin{figure*}[ht!]
\centering
\vspace*{-15mm}
\includegraphics[width=1.0\textwidth]{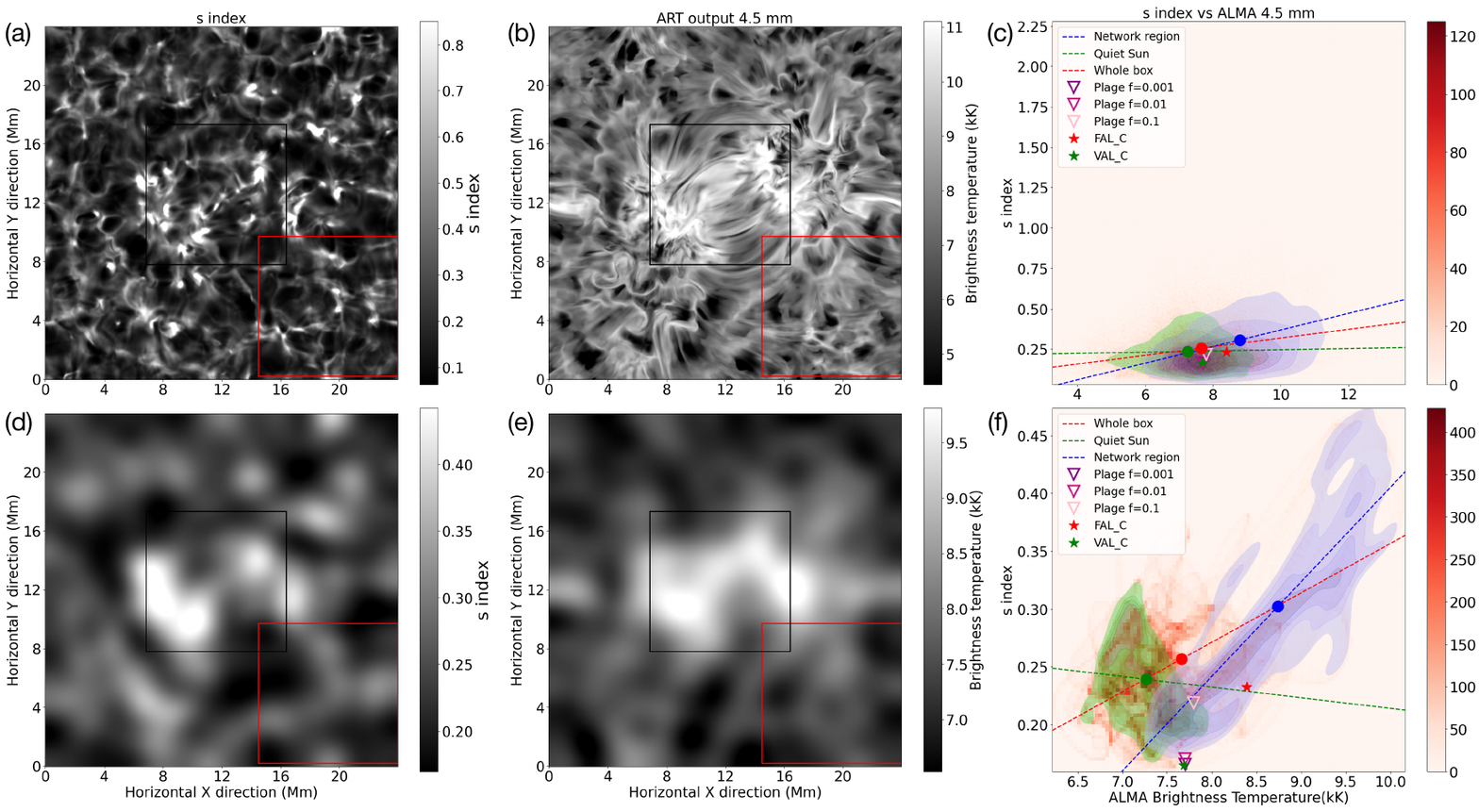}
\vspace*{-20mm}
\caption{Same as Fig.~\ref{fig:CaII_ALMA_3.0mm_comparison_s_index} but for a wavelength of 4.5~mm.}
\label{fig:CaII_ALMA_4.5mm_comparison_s_index}
\end{figure*}

In Figure~\ref{fig:CaII_ALMA_0.3mm_comparison_s_index} the results for 0.3~mm are presented. The wavelength 0.8~mm was of particular interest in the study by \citet{2023A&A...673A.137P}, and thus the corresponding results are presented in Fig.~\ref{fig:CaII_ALMA_0.8mm_comparison_s_index}. In addition to these short wavelengths, the results for a longer wavelength of 4.5~mm in ALMA band 2 are shown in Fig.~\ref{fig:CaII_ALMA_4.5mm_comparison_s_index}. The panel~a in all the four Figs.~\ref{fig:CaII_ALMA_3.0mm_comparison_s_index},~\ref{fig:CaII_ALMA_0.3mm_comparison_s_index},~\ref{fig:CaII_ALMA_0.8mm_comparison_s_index}, and~\ref{fig:CaII_ALMA_4.5mm_comparison_s_index} is the same, but the panels~b show the brightness temperatures at respective wavelengths and panels~d and e are the corresponding degraded maps, using the PSF corresponding to the wavelength. The PSF at 0.3~mm is smaller in width compared to the 3.0~mm PSF by a factor of 0.3/3.0, which can be seen from a visual comparison of Fig.~\ref{fig:CaII_ALMA_3.0mm_comparison_s_index}e and Fig.~\ref{fig:CaII_ALMA_0.3mm_comparison_s_index}e. The 3.0~mm maps appear more blurred, mainly as a result of the wider PSF. The same effect can be seen for the s index maps, which were calculated from the Ca II lines intensity maps after these were all degraded accordingly for each wavelength point.

\begin{figure}
\vspace*{-5mm}
\centering
\includegraphics[width=0.5\textwidth]{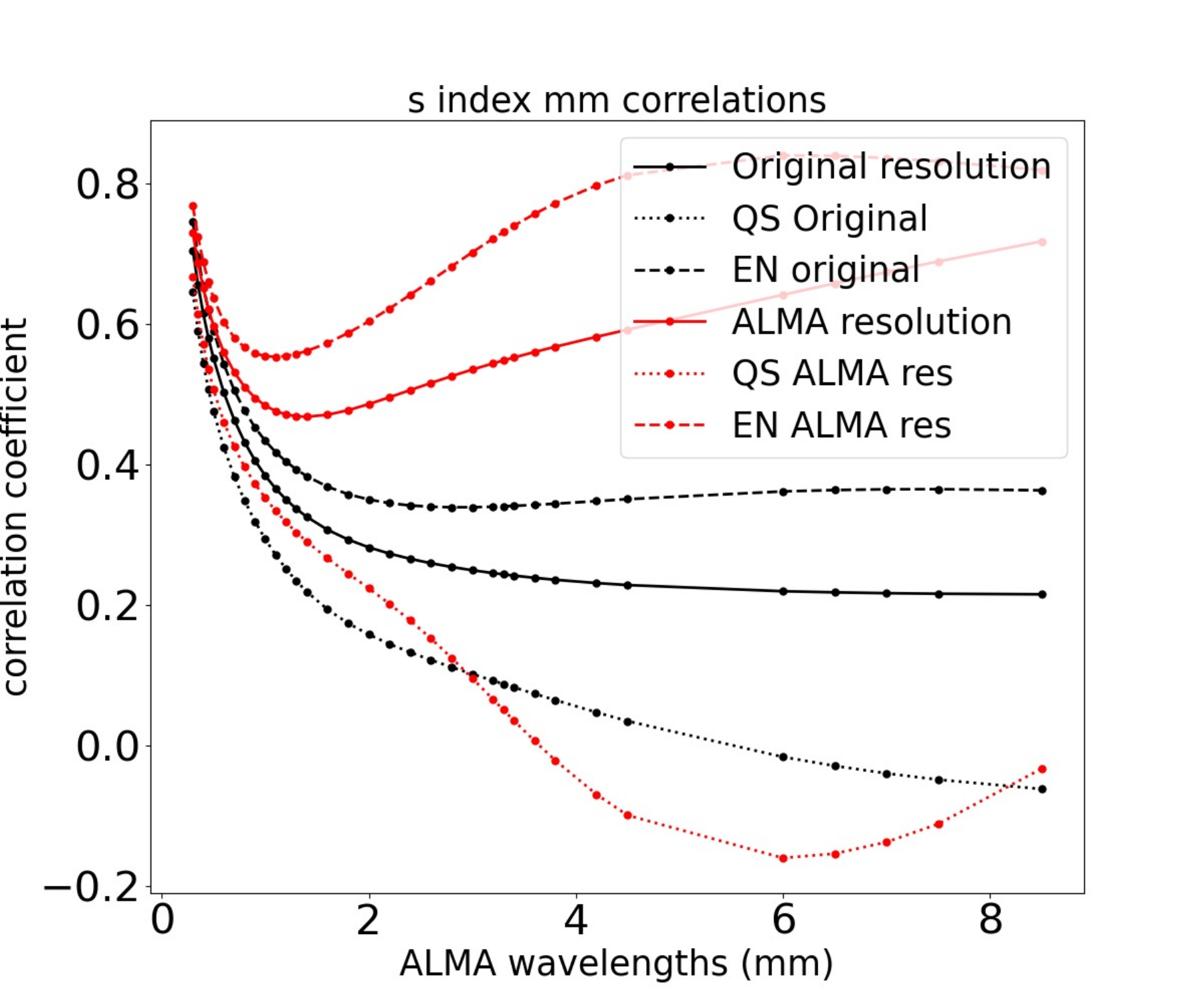}
\caption{Comparison of the calculated Pearson correlation coefficients between s index and mm brightness temperature with two different resolutions}
\label{fig:s_index_ALMA_mm_correlations}
\end{figure}

\begin{figure}
\vspace*{-5mm}
\centering
\includegraphics[width=0.5\textwidth]{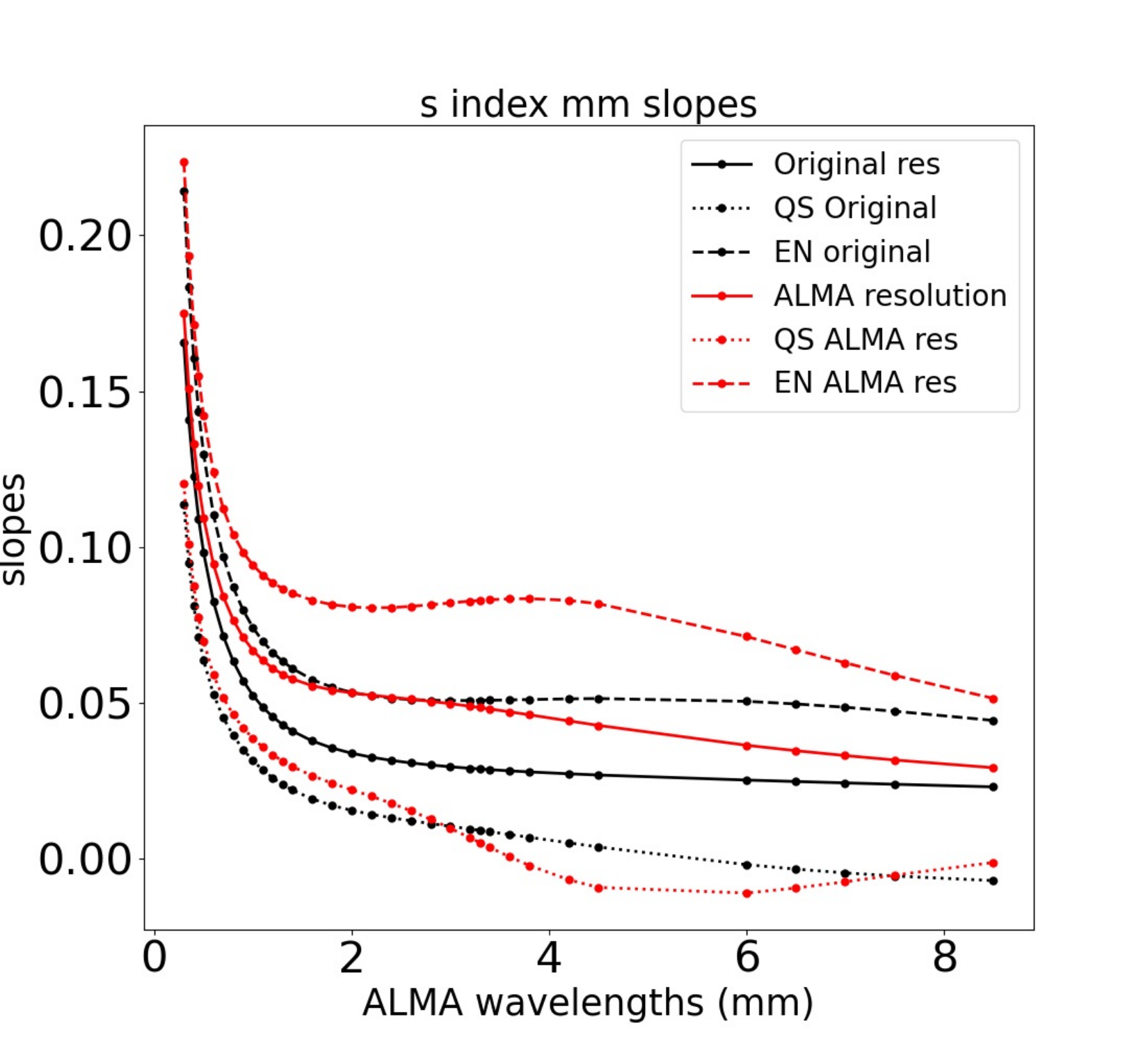}
\caption{Comparison of calculated slopes for scatter plots of s index mm brightness temperatures.}
\label{fig:s_index_ALMA_mm_slopes}
\end{figure}

The calculated Pearson correlation coefficients between the s index and the mm brightness temperatures show that these indices correlate best with the lowest wavelength (see Table~\ref{table:all_three_combined} for the correlation coefficients for 0.3\,mm and 4.5\,mm). The black lines which show the correlation at the original resolution, are lower than the correlations for the degraded resolution as expected. Also, the correlation coefficients are highest for EN and lowest for QS cases in both the resolutions, respectively, similar to the study by \citet{2023A&A...673A.137P}. 

The increase in the correlation coefficients in the case of the whole box and the EN region at the degraded resolution is primarily because of the spatial resolution, namely, the width of the PSF which increases with wavelength. Hence, the structure at spatial scales below the corresponding angular resolution is blurred, which affects the resulting correlation coefficient. With increasing wavelength (and thus decreasing angular resolution) more of this atmospheric fine structure becomes blurred (as seen from a comparison of Figs.~\ref{fig:CaII_ALMA_0.3mm_comparison_s_index} ~and~\ref{fig:CaII_ALMA_4.5mm_comparison_s_index}). However, at the longest wavelengths even structure on relatively large spatial scales is lost. Consequently, the data ranges for these very blurred maps become very narrow, which then results in a high correlation coefficient as seen in Fig.~\ref{fig:s_index_ALMA_mm_correlations}. Most of the original resolution data is very weakly correlated with the Pearson coefficient < 0.3. With degraded resolution, the differences in the nearby pixels get smoothed out, which effectively reduces the corresponding data ranges. 

The slopes calculated for the scatter plots show sudden decay with the wavelengths increasing beyond 1~mm. The slopes are highest at the lowest wavelength and for the wavelengths higher than 1~mm, the slopes are stably decreasing. This is also evident from Figs.~\ref{fig:CaII_ALMA_3.0mm_comparison_s_index}-\ref{fig:CaII_ALMA_4.5mm_comparison_s_index} (panels c,f). For the QS region at degraded resolution at wavelengths longer than 3~mm, the slopes are very close to zero (also observed from  Figs.~\ref{fig:CaII_ALMA_3.0mm_comparison_s_index}c,f,\ref{fig:CaII_ALMA_4.5mm_comparison_s_index}c,f). As the wavelength increases, the QS part in the mm maps and s index map becomes progressively different in terms of structure and the network-internetwork regions become more and more misaligned; this is evident from the green contours being more concentrated in shorter ranges of s index and temperatures. At higher mm wavelengths, the radiation is formed significantly higher up in the atmosphere, whereas the Ca II H\&K fluxes, (the basic constituents of the s index) are formed primarily in the lower atmosphere. Consequently, the scatter plots exhibit correlation coefficients and slopes that approach zero or become negative, indicating a lack of correspondence between the two diagnostics.

\begin{table*}[ht!]
\begin{center}
\npdecimalsign{.}
\nprounddigits{3}

\centering

\scalebox{0.86}{

\begin{tabular}{|c|c|c|c|c|c|c|c|}
\hline
\multicolumn{2}{|c}{}& 
\multicolumn{3}{|c}{Original resolution}& 
\multicolumn{3}{|c|}{Reduced resolution}\\
\cline{2-8}
 & {wavelength}  & {Whole box} &         {QS}    &  {EN} & {Whole box}  &  {QS} &  {EN} \\
\hline
Temperatures &  0.3~mm &        4.722&  4.691&  4.85&   4.722&  4.691&  4.850\\
(kK) &0.8~mm &  5.275    &5.124 &       5.590 & 5.275 & 5.126 & 5.593   \\
&       3.0~mm &        6.968 & 6.621    &7.835 &       6.968 & 6.637 & 7.848   \\
&4.5~mm &       7.661&  7.250&  8.715&  7.661&  7.268&  8.699\\
\hline
\multicolumn{2}{c}{s index} &
\multicolumn{2}{c} {} &
\multicolumn{2}{c|}{} \\                
\hline

Correlation   & 0.3~mm & 0.704  $\pm$ 1.29E-03 & 0.645  $\pm$ 1.27E-03 & 0.744  $\pm$ 1.28E-03 & 0.729  $\pm$ 1.17E-03 & 0.666   $\pm$ 3.22E-03 & 0.767  $\pm$ 2.32E-03 \\
 & 0.8~mm & 0.431   $\pm$ 1.62E-03 & 0.348  $\pm$ 1.64E-03 & 0.476  $\pm$ 1.62E-03 & 0.510  $\pm$ 1.49E-03 & 0.396   $\pm$ 4.49E-03 & 0.566   $\pm$ 2.66E-03 \\
 & 3.0~mm   & 0.249  $\pm$ 1.85E-03 & 0.101  $\pm$ 1.86E-03 & 0.339  $\pm$ 1.85E-03 & 0.534 $\pm$ 1.57E-03 & 0.095  $\pm$ 4.85E-03 & 0.701  $\pm$ 2.19E-03 \\
 & 4.5~mm & 0.223  $\pm$ 2.01E-03 & 0.034 $\pm$ 1.97E-03 & 0.350  $\pm$ 1.99E-03 & 0.591  $\pm$ 1.59E-03 & -0.099 $\pm$ 6.38E-03 & 0.811  $\pm$ 1.49E-03 \\
\hline
Slopes & 0.3~mm & 0.165  $\pm$ 3.31E-04 & 0.114  $\pm$ 6.75E-04 & 0.214  $\pm$ 9.61E-04 & 0.175  $\pm$ 3.26E-04 & 0.120   $\pm$ 6.74E-04 & 0.224  $\pm$ 9.34E-04 \\
 & 0.8~mm & 0.063 $\pm$ 2.63E-04 & 0.039 $\pm$ 5.32E-04 & 0.087 $\pm$ 8.04E-04 & 0.077 $\pm$ 2.56E-04 & 0.046  $\pm$ 5.36E-04 & 0.104   $\pm$ 7.58E-04 \\
 & 3.0~mm   & 0.029 $\pm$ 2.28E-04 & 0.010 $\pm$ 5.11E-04 & 0.051 $\pm$ 7.05E-04 & 0.049 $\pm$ 1.56E-04 & 0.009 $\pm$ 5.11E-04 & 0.082 $\pm$ 4.17E-04 \\
 & 4.5~mm & 0.027 $\pm$ 2.28E-04 & 0.004 $\pm$ 5.50E-04 & 0.051 $\pm$ 6.87E-04 & 0.043 $\pm$ 1.16E-04 & -0.009 $\pm$ 4.59E-04 & 0.082 $\pm$ 2.95E-04 \\
  \hline
\multicolumn{2}{c}{IRT index} &
\multicolumn{2}{c} {} &
\multicolumn{2}{c|}{ } \\                                               
\hline

Correlation   & 0.3~mm & 0.535  $\pm$ 1.77E-03 & 0.428   $\pm$ 1.72E-03 & 0.621  $\pm$ 1.74E-03 & 0.559  $\pm$ 1.64E-03 & 0.446   $\pm$ 4.20E-03 & 0.645  $\pm$ 3.29E-03 \\
 & 0.8~mm & 0.307  $\pm$ 1.91E-03 & 0.212  $\pm$ 1.92E-03 & 0.383  $\pm$ 1.92E-03 & 0.382  $\pm$ 1.83E-03 & 0.253   $\pm$ 4.87E-03 & 0.472  $\pm$ 3.45E-03 \\
 & 3.0~mm   & 0.176  $\pm$ 1.96E-03 & 0.068 $\pm$ 1.97E-03 & 0.289 $\pm$ 1.93E-03 & 0.467  $\pm$ 1.69E-03 & 0.090  $\pm$ 4.40E-03 & 0.660  $\pm$ 2.16E-03 \\
 & 4.5~mm & 0.164  $\pm$ 2.02E-03 & 0.019 $\pm$ 1.99E-03 & 0.296  $\pm$ 2.04E-03 & 0.514  $\pm$ 1.81E-03 & -0.139  $\pm$ 5.54E-03 & 0.763  $\pm$ 1.53E-03 \\

\hline
Slopes & 0.3~mm & 0.033 $\pm$ 1.04E-04 & 0.022 $\pm$ 2.36E-04 & 0.043 $\pm$ 2.76E-04 & 0.035 $\pm$ 1.03E-04 & 0.024 $\pm$ 2.38E-04 & 0.046 $\pm$ 2.72E-04 \\
 & 0.8~mm & 0.012 $\pm$ 7.29E-05 & 0.007 $\pm$ 1.64E-04 & 0.017 $\pm$ 2.07E-04 & 0.015 $\pm$ 7.09E-05 & 0.008 $\pm$ 1.63E-04 & 0.021 $\pm$ 1.96E-04 \\
 & 3.0~mm & 0.005 $\pm$ 6.09E-05 & 0.002 $\pm$ 1.52E-04 & 0.011 $\pm$ 1.75E-04 & 0.010 $\pm$ 3.79E-05 & 0.002 $\pm$ 1.34E-04 & 0.017 $\pm$ 9.65E-05 \\
 & 4.5~mm & 0.005 $\pm$ 6.07E-05 & 0.001 $\pm$ 1.63E-04 & 0.011 $\pm$ 1.72E-04 & 0.008 $\pm$ 2.75E-05 & -0.003 $\pm$ 1.15E-04 &  0.016 $\pm$  6.90E-05 \\
\hline
\end{tabular}}
\caption{Brightness temperatures in kK, correlation coefficients and slopes for the three pixel sets (whole box, QS and EN) each for two resolutions (original and degraded) for the s index and IRT index.}
\label{table:all_three_combined}
\end{center}
\end{table*}

\subsection{The IRT index}
\begin{figure*}[ht!]
\centering
\vspace*{-15mm}
\includegraphics[width=1.0\textwidth]{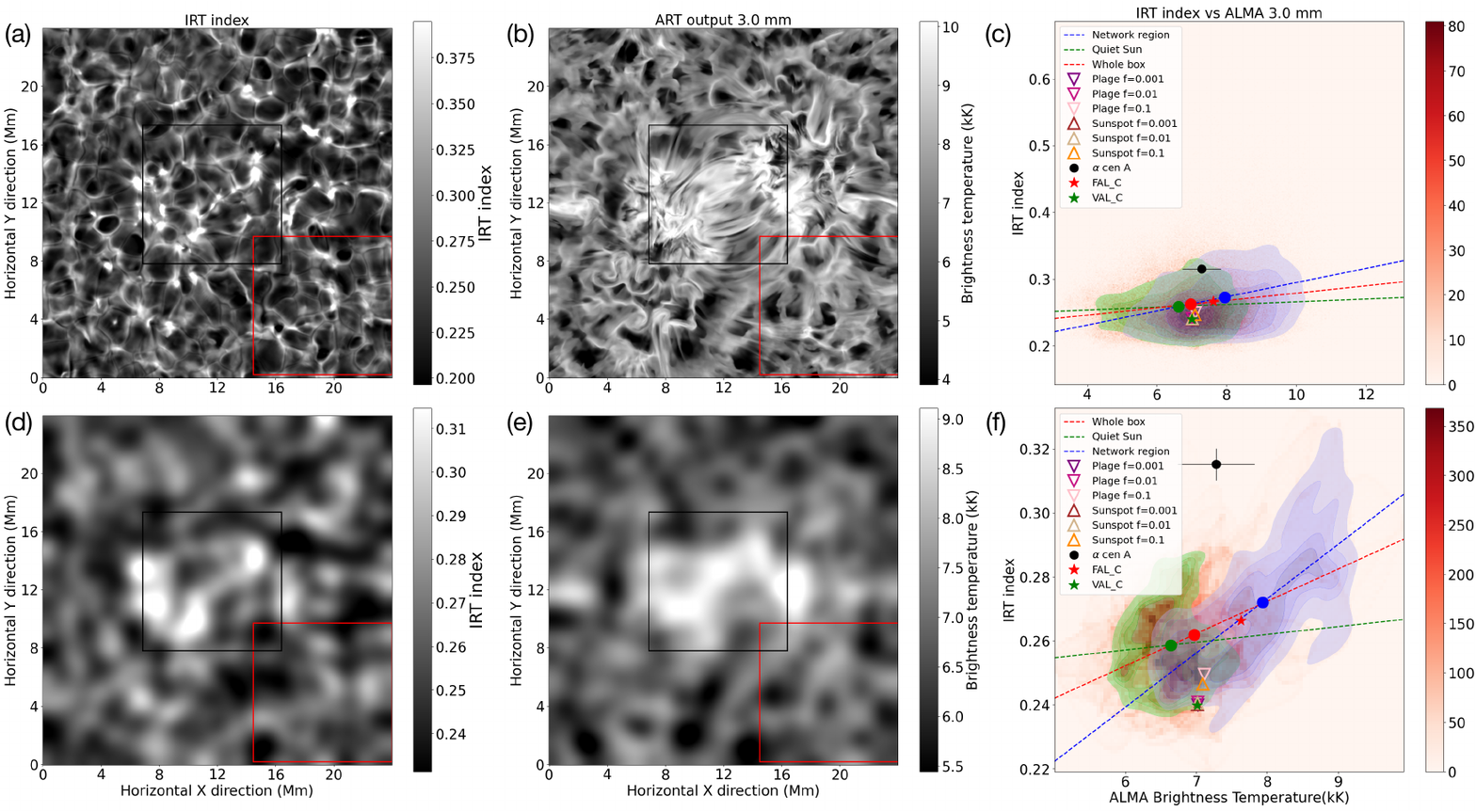}
\vspace*{-20mm}
\caption{Same as Fig.~\ref{fig:CaII_ALMA_3.0mm_comparison_s_index} but for IRT index, and the black circle with error bars is an observational data point for G2V type star $\alpha$ Cen A with calculated IRT index of 0.315.}
\label{fig:CaII_ALMA_3.0mm_comparison_IRT_index}
\end{figure*}

\begin{figure*}[ht!]
\centering
\vspace*{-15mm}
\includegraphics[width=1.0\textwidth]{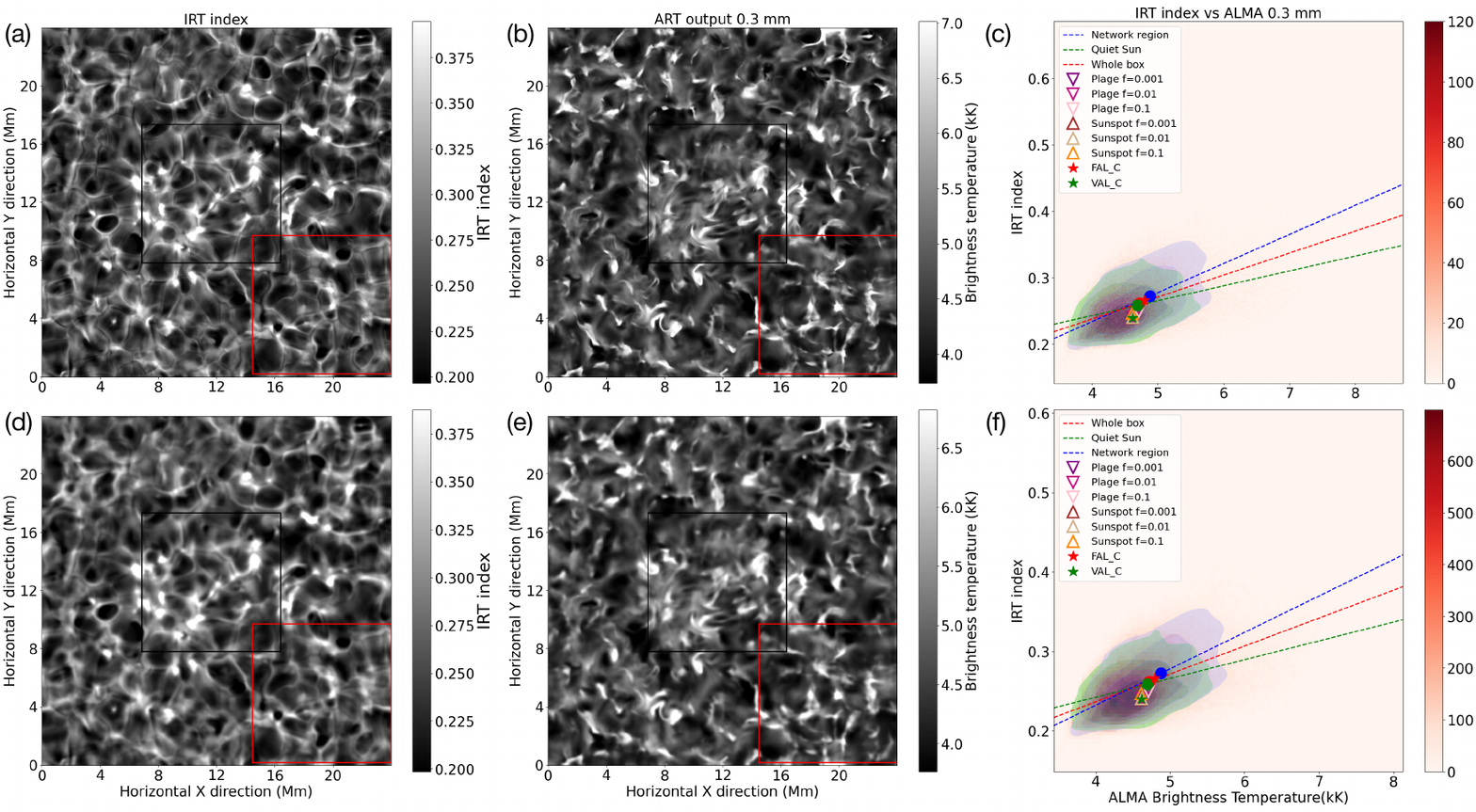}
\vspace*{-20mm}
\caption{Same as Fig.~\ref{fig:CaII_ALMA_3.0mm_comparison_IRT_index} but for a wavelength of 0.3~mm.}
\label{fig:CaII_ALMA_0.3mm_comparison_IRT_index}
\end{figure*}

\begin{figure*}[ht!]
\centering
\vspace*{-15mm}
\includegraphics[width=1.0\textwidth]{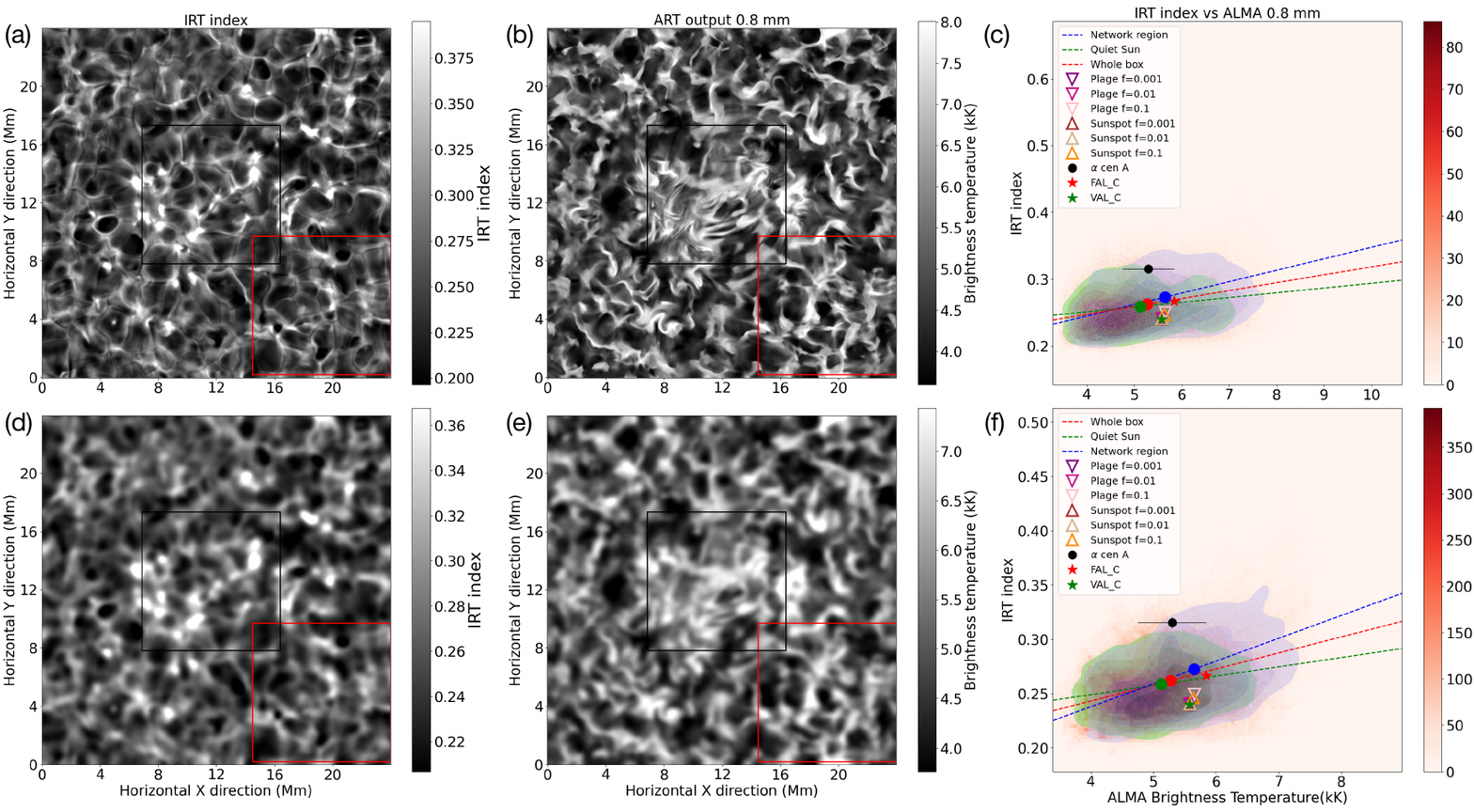}
\vspace*{-20mm}
\caption{Same as Fig.~\ref{fig:CaII_ALMA_3.0mm_comparison_IRT_index} but for a wavelength of 0.8~mm.}
\label{fig:CaII_ALMA_0.8mm_comparison_IRT_index}
\end{figure*}

\begin{figure*}[ht!]
\centering
\vspace*{-15mm}
\includegraphics[width=1.0\textwidth]{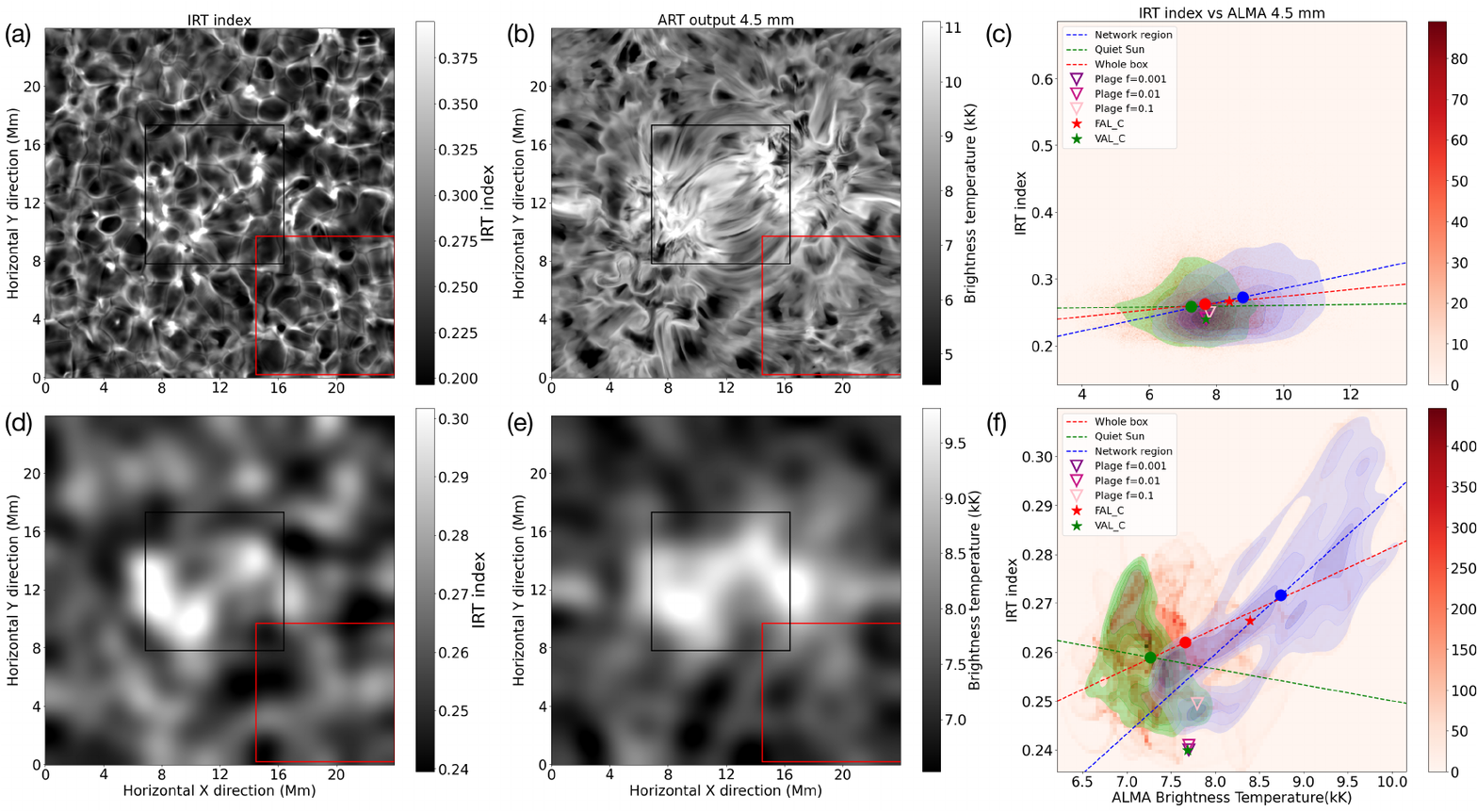}
\vspace*{-20mm}
\caption{Same as Fig.~\ref{fig:CaII_ALMA_3.0mm_comparison_IRT_index} but for a wavelength of 4.5~mm.}
\label{fig:CaII_ALMA_4.5mm_comparison_IRT_index}
\end{figure*}

An analysis similar to the one presented for the s index was also done for the Ca infrared triplet (IRT) lines. The integrated intensities for the Ca IRT lines over the rectangular filters described in Fig.~\ref{fig:15k_vs_789grid} are shown in Fig.~\ref{fig:Ca_II_maps}~(d,e~and~f). In comparison to the K and H lines, the IRT lines are formed at slightly lower heights in the lower chromosphere. Structures showing reverse granulation features are seen and the intensities are slightly higher at the footpoints of the loops, in comparison to the internetwork regions where shock fronts are observed. As shown by \citet{2024A&A...682A..11M}, the synthetic Ca~II~854.2~nm data from Bifrost shows narrower profiles than the observed solar data derived with the CRISP instrument at the Swedish 1-m Solar Telescope (SST). This means that the Bifrost model in effect underestimates the IRT index. The IRT index at the original resolution calculated using the fluxes and the continuum (as per Eq.~\ref{eq:IRT}) are shown in  Fig.~\ref{fig:CaII_ALMA_3.0mm_comparison_IRT_index}~a, which shows lower layer in the atmosphere in comparison to Fig.~\ref{fig:CaII_ALMA_3.0mm_comparison_s_index}a. Similar to the s index case, the original resolution maps for the IRT index and mm brightness temperature are shown in panels a and b and the degraded resolution maps are shown in panels d and e, respectively. The scatter contour plots are plotted for both resolutions in panels c and f,  respectively, similar to the treatment in the previous section. Also, similar plots for a wavelength of 0.3~mm showing highest correlation coefficient, 0.8~mm, being of particular interest in \citet{2023A&A...673A.137P},  and 4.5~mm for illustrating wavelength formed at higher heights are also shown respectively in Figs.~\ref{fig:CaII_ALMA_0.3mm_comparison_IRT_index},~\ref{fig:CaII_ALMA_0.8mm_comparison_IRT_index}, and~\ref{fig:CaII_ALMA_4.5mm_comparison_IRT_index}. The mean values of the IRT index for the whole box, QS and EN regions are 0.262, 0.259 and 0.273, shown by filled circles of the red, green and blue colour, respectively.

\begin{figure}
\vspace*{-5mm}
\centering
\includegraphics[width=0.5\textwidth]{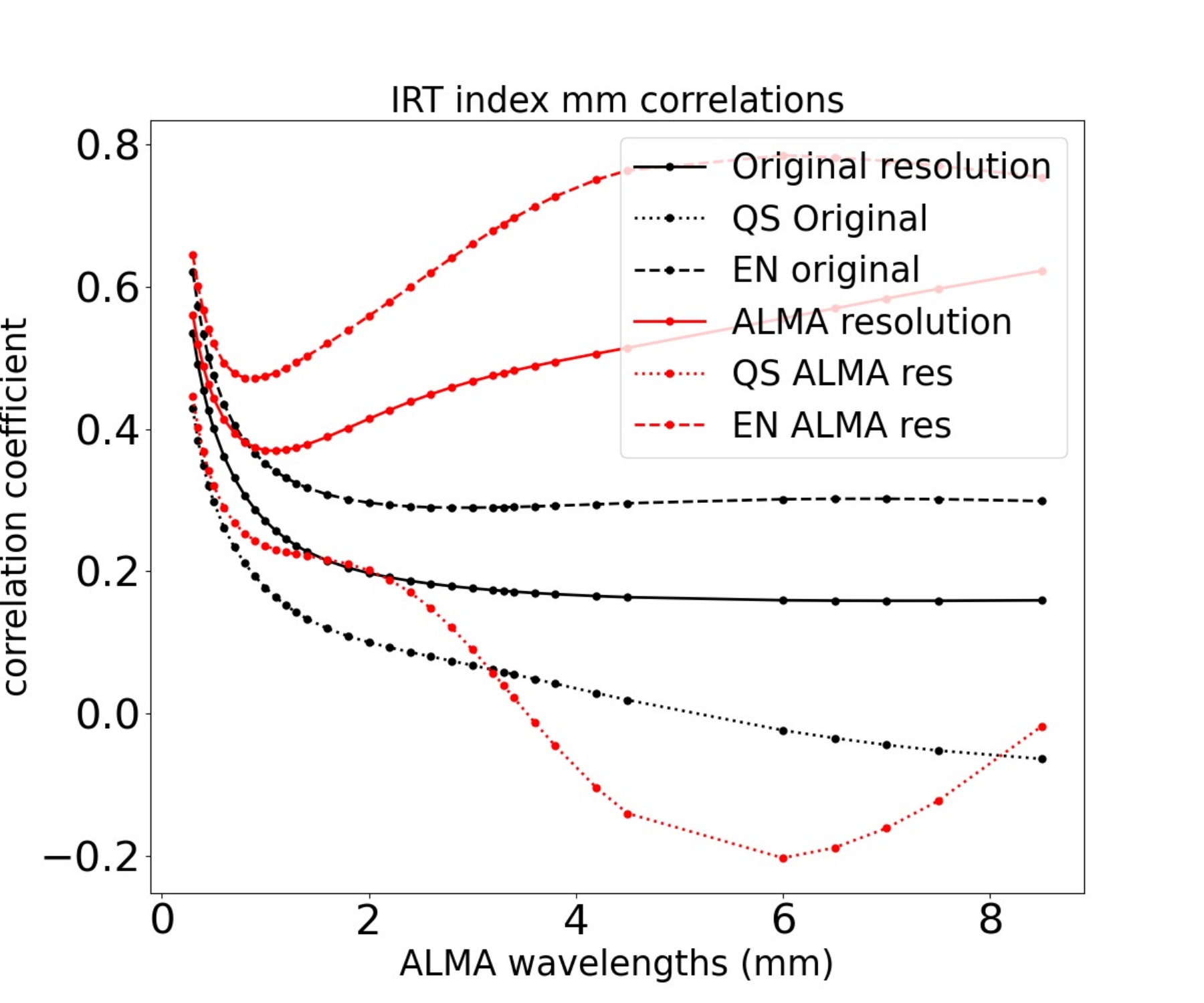}
\caption{Comparison of the calculated Pearson correlation coefficients between IRT index and mm brightness temperature with two different resolutions}
\label{fig:IRT_index_ALMA_mm_correlations}
\end{figure}

\begin{figure}
\vspace*{-5mm}
\centering
\includegraphics[width=0.5\textwidth]{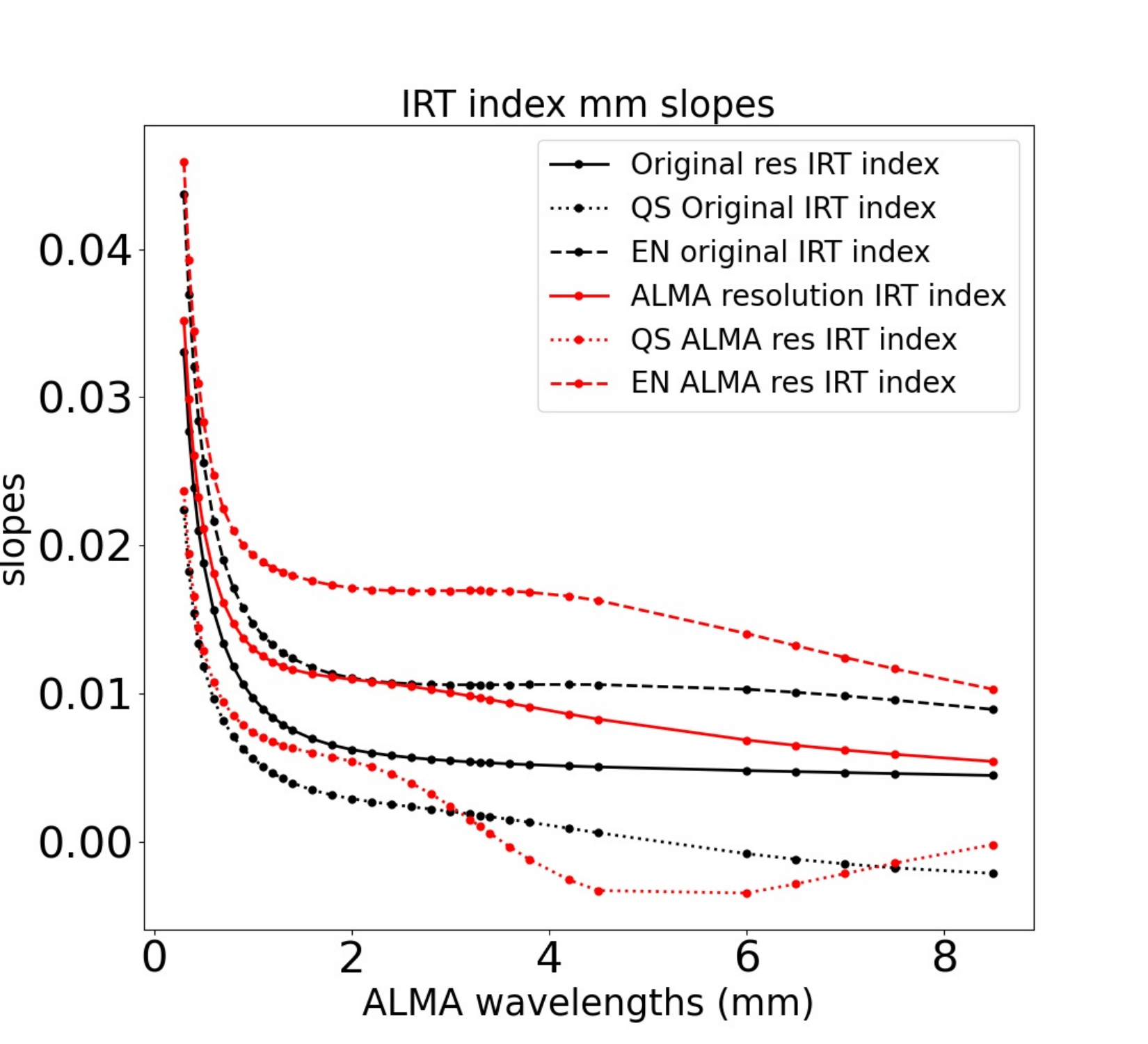}
\caption{Comparison of the calculated slopes for the scatter plots for IRT index and mm brightness temperatures.}
\label{fig:IRT_index_ALMA_mm_slopes}
\end{figure}

Figures~\ref{fig:IRT_index_ALMA_mm_correlations} and \ref{fig:IRT_index_ALMA_mm_slopes} reveal that the correlation coefficients and slopes peak at the lowest wavelengths, similar to the correlation coefficients and slopes derived for the s index. However, the slopes for the IRT index are an order of magnitude lower, evident from the fact that the range of the calculated IRT index is an order of magnitude lower than the calculated s index. The correlations are slightly lower as well, which can be attributed to the IRT index plane representing a slightly lower atmospheric region compared to the s index plane. This distinction is evident in Fig.\ref{fig:CaII_ALMA_3.0mm_comparison_IRT_index}a and Fig.\ref{fig:CaII_ALMA_3.0mm_comparison_s_index}a. Moreover, Fig.\ref{fig:CaII_ALMA_0.3mm_comparison_s_index}b and Fig.\ref{fig:CaII_ALMA_0.3mm_comparison_IRT_index}b demonstrate that the 0.3~mm map bears more similarity to the s index than to the IRT index. Thus, the correlation coefficient for the IRT index is lower than that for the s index.

It is important to note that the general trends in the correlation coefficients and slopes are very similar for both the s index and the IRT index. In the original resolution case, the correlation coefficients decrease from 0.3~mm to 1.0~mm, but the decrease is not significant for longer wavelengths. In the degraded resolution case, however, the correlation coefficients notably increase at longer wavelengths due to the larger corresponding PSFs. The slopes also exhibit a drastic decrease between 0.3~mm and 1.0~mm and this trend is observed in both the original resolution and degraded resolution cases. With further increases in wavelength, the decrease in slopes becomes less significant. Additionally, Table~\ref{table:all_three_combined} demonstrates that for degraded resolution, the slopes are consistently higher compared to the corresponding original resolution case, but follow the same pattern of a sudden decrease at smaller wavelengths and a gradual decrease at longer wavelengths.

Figures~\ref{fig:CaII_ALMA_0.3mm_comparison_s_index}c\&f~and~\ref{fig:CaII_ALMA_0.3mm_comparison_IRT_index}c\&f demonstrate larger slopes at lower wavelengths, indicated by the strong correspondence between the linear fits for the whole box, QS, and EN regions.  Conversely, in Figs.~\ref{fig:CaII_ALMA_4.5mm_comparison_s_index}c\&f~and~\ref{fig:CaII_ALMA_4.5mm_comparison_IRT_index}c\&f, the lines appear almost horizontal due to the lack of corresponding distributions at longer wavelengths. For the QS, the slopes become more negative as the wavelength increases. This trend is also observed in the correlation coefficients for QS at the degraded resolution, as shown in Table~\ref{table:all_three_combined}. In Figs.~\ref{fig:s_index_ALMA_mm_correlations}~and~\ref{fig:IRT_index_ALMA_mm_correlations}, the correlations decrease and become negative for wavelengths longer than 3~mm. This is explained by the respective patterns of shock fronts in Figs.~\ref{fig:CaII_ALMA_3.0mm_comparison_s_index},~\ref{fig:CaII_ALMA_4.5mm_comparison_s_index},~~\ref{fig:CaII_ALMA_3.0mm_comparison_IRT_index}, and~\ref{fig:CaII_ALMA_4.5mm_comparison_IRT_index} (panels d and e). The QS region shows shock patterns that do not correspond well to each other, resulting in negative slopes in panel f, as observed in Table~\ref{table:all_three_combined}. Conversely, the EN region exhibits strong correspondence, reflected in the contours with highly positive slopes and higher correlation coefficients compared to the original resolution case. The calculated uncertainties, using bootstrapping in the case of correlation coefficients and by using the residues in the case of slopes, are 2 to 3 orders of magnitude smaller than the values of correlation coefficients and slopes as seen in Table~\ref{table:all_three_combined}, making these results reliable.

\section{Discussion}
\label{Sec:Discussion} 

\subsection{Formation heights}
In Table~\ref{tab:formation heights}, the formation heights of the line feature from the solar photosphere ($\tau_{500}=1$ surface) are listed. \cite{2018A&A...611A..62B}, constrained the formation height range using SST/CHROMIS observations and different models for Ca II absorption-emission peaks generated from Bifrost atmospheres. The average formation heights for the Ca II K line are provided with respect to the height where the average optical depth is unity, whereas the Ca~II~H formation height range is provided with respect to the K line. Our results also show very similar formation heights. The average formation height for the line cores of K and H are found to be 1.73~Mm and 1.62~Mm, respectively, which are within the range described in \citet{2018A&A...611A..62B}.  As seen in Fig.~\ref{fig:Ca_II_maps}b,c, the K line core map shows fewer shock fronts and internetwork features, which are formed at slightly lower heights, than the H line. Further, the formation heights of the IRT line cores are lower than those of the H\&K lines, namely 0.95~Mm, 1.17~Mm, and 1.11~Mm for 849.8~nm, 854.2~nm, and 866.2~nm,  respectively. The infrared triplet line cores are formed at lower heights than the H\&K lines \citep[as seen in Fig.~15 in ][]{2018A&A...611A..62B}, which is very evident from the integrated intensity maps in Fig.~\ref{fig:Ca_II_maps}.

\subsection{Comparison with semi-empirical models}
In this study, the Ca II indices and the brightness temperatures~$T_b$ at mm wavelengths are compared for the semi-empirical model atmospheres VAL~C and FAL~C. Figure~6 in \citet{2023A&A...673A.137P} illustrates that the VAL~C model consistently exhibits lower brightness temperature compared to the FAL~C model. The publicly available 3D simulations of the solar atmosphere by \citet{2016A&A...585A...4C} reveal that the Mg~II h \& k lines are narrow in comparison to the observations made by IRIS \citep{2019ARA&A..57..189C}. The Ca II line profiles for the VAL~C and FAL~C models are seen to be broader than the line profiles calculated with the Bifrost model, suggesting that the models lack sufficient mass at higher temperatures in the upper chromosphere \citep{2023A&A...673A.137P}. Consequently, these semi-empirical models of the quiet Sun exhibit slightly broader profiles, lower integrated intensities, and consequently lower Ca II indices. However, the Ca II index values for the semi-empirical models lie within the range of the EN and QS distributions on the Ca II indices vs temperature planes, indicating an agreement between the data from the 3D close-to-realistic model and semi-empirical models. On the other hand, the estimated average s index for the Sun is comparable, near 0.17 as found in \citet{2017ApJ...835...25E}, which is close to the observational value of s index for the sun-like star $\alpha$~Cen~A. The calculated s index from the 3D model and the 1D semiempirical models slightly overestimate the s index.

\subsection{Other Ca II indices}
In this study only the s index and IRT index are considered for comparison with the mm brightness temperatures and not the $R_{HK}, R'_{HK}, R_{IRT},  R'_{IRT}$ indicators generated from Ca II lines. The latter indices are defined in \citet{1984ApJ...279..763N} which are derived from \citet{1982A&A...107...31M}, using colour-dependent conversion factor $C_{cf}(B-V)$ which converts the s index to $R_{HK}$ index as $R_{HK} = C_{cf}(B-V) \times s.$ The $R_{HK}$ index is fundamentally the Ca II H\& K flux corrected for photospheric radiation, to make it a better chromospheric indicator. Further $R_{HK}$ index is normalised to bolometric luminosity to generate $R'_{HK}$ index. When comparing the calcium activity indicators with mm continuum intensities generated from a 3D solar model atmosphere, $B-V$ and the bolometric luminosity, both would be constants. Hence, while looking at the relations, it does not change the correlation coefficients and the slopes discussed in the paper. Furthermore, based on the stellar sample studied, the exact polynomial used to calculate the colour-dependent conversion factor $C_{cf}(B-V)$ is different \citep[for example, ][ and references therein.]{2022A&A...664L...9M, 2018A&A...616A.108B, 2014MNRAS.444.3517M, 1984ApJ...279..763N}. 

\subsection{Bandpasses for H\&K lines integrated intensities}
\citet{2021ApJ...914...21S} use rectangular filters instead of triangular filters to calculate the integrated intensities for calculating the s index.  When comparing the correlations and slopes for the comparison of s index and mm wavelength brightness temperatures, correlations are highest at the lowest wavelengths for both the rectangular and the triangular filter integrated intensities. The only noticeable difference is that the rectangular filter gives slightly higher integrated intensity values in the integrated signal as the parts in the line which are slightly away from the core are given more weight in comparison to the triangular filter. Hence, using the rectangular filter would contaminate the H \& K integrated intensities with information from the lower chromosphere than the height at which the line cores are formed. For the purposes of our study, given that this would be beneficial for stellar astrophysical studies in the future, we used the traditional definition for s index with triangular filters, as defined by \citet{1978PASP...90..267V}.

\subsection{Implication for solar-like stars}
\label{sec:sun-like-star}

A direct application to stellar observations is to estimate the stellar activity level, in particular for Sun-like stars for instance used in the studies by \citet{2016A&A...594A.109L} and \citet{2021A&A...655A.113M, 2022A&A...664L...9M}. The data point for a sun-like star in the Ca II indices--mm brightness temperature plane as derived from adequate observations then provides constraints on the properties of the observed stellar atmosphere. This approach is illustrated by comparing the data points for the semi-empirical 1D models FAL~C and VAL~C to the averages for the whole box, EN, and QS data points on the Ca II indices--mm brightness temperature plane in Figs.~\ref{fig:CaII_ALMA_3.0mm_comparison_s_index}~(c and f) and \ref{fig:CaII_ALMA_3.0mm_comparison_IRT_index}~(c and f). 

For the disk-integrated signal, the centre-to-limb variation for the Ca lines and ALMA full disk data should be taken into account \citep[see e.g.,][and references therein]{2022FrASS...9.1320A, 2022A&A...661L...4A, 2018A&A...619L...6N}. As the Ca II line cores form in the chromosphere and the wings in the photosphere, the limb effects would vary (brightening or darkening depending on the wavelength) when going from the core to the wing. In a recent study based on SST observations, \citet{2023A&A...671A.130P} concluded that chromospheric lines, like H$\alpha$, Ca~II~H~\&~K, and Ca~II~8542, might exhibit a blueshift towards the limb due to the chromospheric canopies, but the effect on the indices is non-trivial, as the line core is more affected by the limb effects than the nearby continuum. Further, the authors show that the line core widths would decrease when going from the centre to the limb. This is in line with the previous works, where \citet{1966ApNr...10..101E} stated that the centre-to-limb variation is a function of the turbulent velocities in the plasma. Further systematic studies need to be conducted to understand the limb effects on the correlations between the diagnostics and the implications for disk-integrated stellar observations.

The brightness temperature values at the wavelengths of 0.8~mm and 3~mm for $\alpha$~Cen~A, which are obtained from ALMA observations \citep[see][and references therein]{2021A&A...655A.113M}, are in line with the values derived from ALMA observations of the Sun and also with those based on the simulations presented in this study, listed in Table~\ref{table:all_three_combined}. In contrast, the calculated Ca II s index for $\alpha$~Cen~A taken from \citet{1996AJ....111..439H}, is 0.165,  although varying in time between 0.157 and 0.167, and thus in line with the values calculated for the FAL~C and VAL~C models (see Figs.~\ref{fig:CaII_ALMA_3.0mm_comparison_s_index}~c,f  and \ref{fig:CaII_ALMA_0.8mm_comparison_s_index}~c,f). In comparison, the corresponding average value derived from the synthetic observables for the Bifrost simulation is 0.256 and thus substantially higher than the value for FAL~C and VAL~C and the value for $\alpha$~Cen~A. The IRT index for $\alpha$~Cen~A is calculated to be 0.315 using the observational data described in Sect.~\ref{sec:Alf_cen_A_data} and thus significantly higher than the average IRT index of 0.262 for the simulated data from the Bifrost model atmosphere, which causes the respective $\alpha$~Cen~A data point to be placed away from the linear fit for the solar simulation shown in Figs.~\ref{fig:CaII_ALMA_0.3mm_comparison_IRT_index} (c and f) and \ref{fig:CaII_ALMA_0.8mm_comparison_IRT_index} (c and f). The black circles show the error bars in these figures.

The wider Ca II line profiles observed for $\alpha$~Cen~A compared to the Sun may be attributed to differences in atmospheric structure and activity levels, but a thorough investigation is required to draw any firm conclusions. Additionally, the solar observations are limited to a small, spatially resolved region on the Sun, while the $\alpha$~Cen~A data are integrated across the entire (spatially unresolved) stellar disk. The used model has a large filling factor of bright internetwork, which might make the average of the distributions of limited use for comparison with full-disc integrated observations of the Sun-like star. Nonetheless, the semi-empirical model shows similar values of indices with the $\alpha$~Cen~A data. In \citet{2018A&A...611A..62B} the formation of the Ca~II lines from the same Bifrost model atmosphere along with modified model atmospheres including more physical effects such as non-equilibrium ionisation of He, and instantaneous LTE treatment, are presented and the Ca II line profiles are narrower than the compared solar observational line profiles. They conclude that there is a missing component in all models that explains the wider observed profiles. Two potential explanations for this discrepancy are the absence of motions at scales smaller than the photon mean free path, commonly referred to as microturbulence and insufficiently strong heating processes in the lower chromosphere.

\subsection{Comparison with different activity levels}
\begin{figure}
\centering
\includegraphics[width=0.5\textwidth]{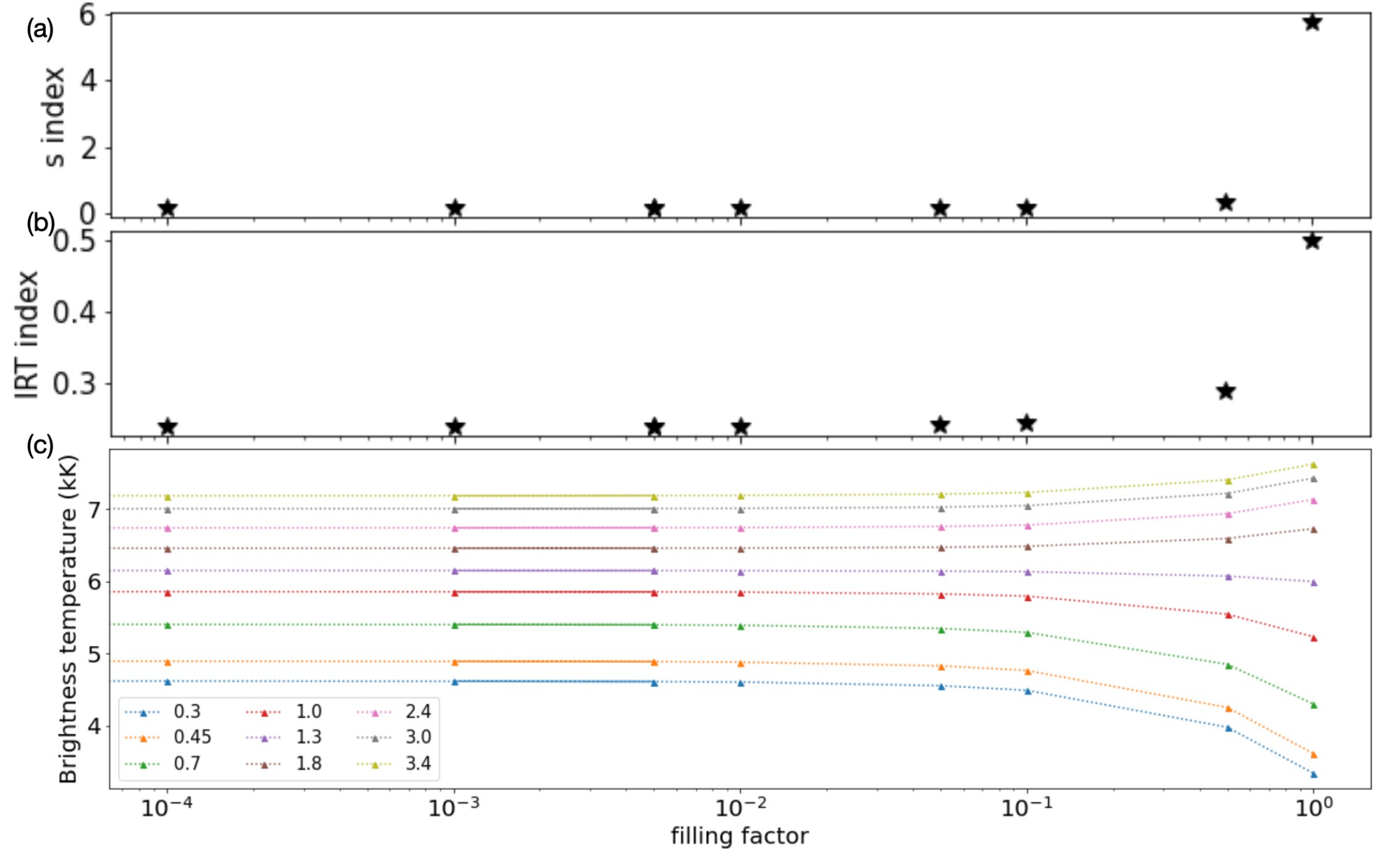}
\vspace{-0.2mm}
\includegraphics[width=0.49\textwidth]{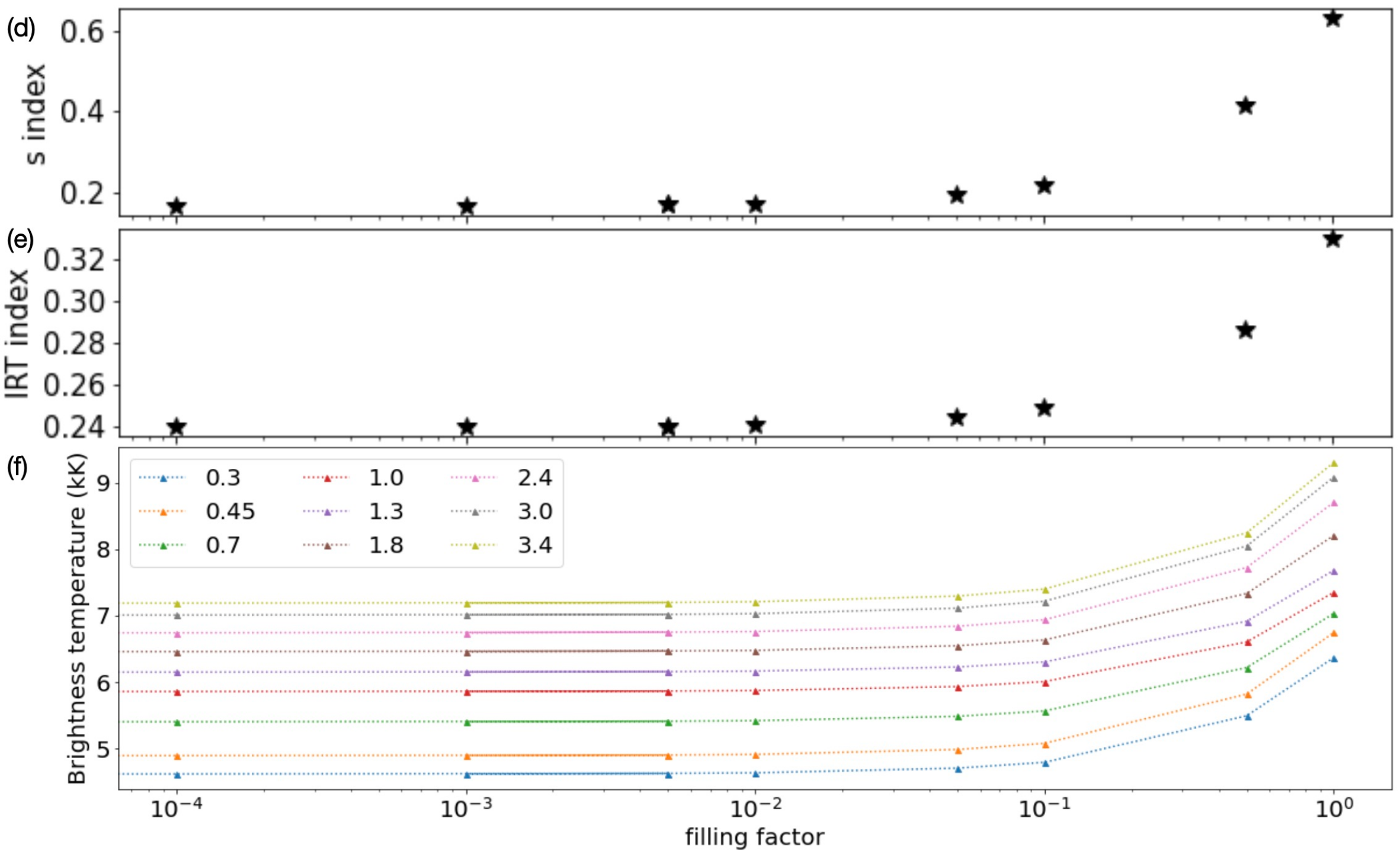}
\caption{s index (panels a and d), IRT index (panels b and e), and the brightness temperatures at the mm wavelengths shown in the legend (panels c and f) as a function of filling factors. The filling factors are considered to be 0.0001, 0.005, 0.001, 0.005, 0.01, 0.05, 0.1, 0.5, and 1.0. The upper box shows plots for the sunspot model by \citet{1994ASIC..433..169S} and the lower box, for the plage model by \citet{2009ApJ...707..482F}.}
\vspace{-5mm}
\label{fig:filling_factors}
\end{figure}

The correlation between the Ca activity indices and the mm brightness temperature can be used for the study of solar-like stars as demonstrated in Sect.~\ref{sec:sun-like-star}. As active region simulations are not immediately available, this study focused on simulated EN and QS regions. The impact of other types of regions on (synthetic) observables and resulting activity indicators as they would be derived from unresolved stellar observations is here explored by combining the radiative transfer results from 1D models with different filling factors, alternatively referred to as the fractional area \citep[see, e.g.,][]{2019arXiv190401133A, 2009JGRA..114.7104B}. For this purpose, the quiet Sun model VAL~C  \citep{1981ApJS...45..635V}, the plage model H by \citet{2009ApJ...707..482F}, and the sunspot umbra model by \citet{1994ASIC..433..169S} as provided by Loukitcheva (priv. comm.) are used. The umbra model was chosen according to the detailed study by \citet{2022FrASS...925368L} that compares various models to recent (interferometric) ALMA sunspot observations \citep[see also][]{2014A&A...561A.133L}. It should be noted that these observations exhibit brightness temperatures in sunspot umbrae in band~6 (1.3\,mm) down to $\sim 5000$\,K, which are significantly lower than in quiet Sun regions \citep[$\sim 5900$\,K, ][]{2017SoPh..292...88W}, while temperatures of up to $\sim 8000$\,K are found in the penumbra. Even at the lower resolution of ALMA Total Power (TP) single-dish observations, brightness temperatures of 5700\,K are observed in the umbra \citep{2017ApJ...850...35L}. In band~3 (3\,mm), the brightness temperatures in the penumbra also reach values of $\sim 8000$\,K while the quiet Sun has on average a brightness temperature of 7300\,K \citep{2017SoPh..292...88W}. In the umbra, however, the brightness temperature first drops slightly but then increases in the innermost parts to about $\sim 8500$\,K and, thus, beyond the penumbral value. This finding has been made possible by the unprecedented spatial resolution achieved with ALMA and is yet to be explained  \citep{2022FrASS...925368L}. In conclusion, in comparison to the average Sun, sunspot umbrae appear as cooler features in band~6 but as slight temperature enhancements in Band~3 \citep{2017ApJ...850...35L}.

The resulting brightness temperatures and activity indicators for combinations of these components for varying the filling factors are shown in Fig.~\ref{fig:filling_factors}. As the sunspot coverage increases, the calcium indices increase, and so do the brightness temperatures for mm wavelengths greater than 1.5~mm. For wavelengths shorter than 1.5~mm, the combined brightness temperatures decrease because sunspot umbrae appear darker at those wavelengths. With increasing plage filling factors, all the calcium indices and the brightness temperatures increase as plage regions are brighter at all the heights from where the mm continuum emerges at the considered wavelength range. Only values for the plage filling factor of more than a few percent result in a notable increase in the continuum brightness temperature for all wavelengths considered and also for the IRT index and the s index. For the Sun, realistic values for the plage filling factor vary from 0 at solar minimum to 0.04 - 0.07 at solar maximum, whereas the filling factor of sunspots ranges from 0 at solar minimum to 0.002 - 0.005 during solar maximum \citep{2022ApJ...937...84P}. Consequently, as illustrated in Fig.~\ref{fig:filling_factors}, the resulting change in average continuum brightness temperature for all considered wavelengths and the activity indices remains moderate for the Sun. The data points for the combined 1D atmospheres with different filling factors are plotted in the Ca II index-mm brightness temperature planes in Figs.~\ref{fig:CaII_ALMA_0.3mm_comparison_s_index},~\ref{fig:CaII_ALMA_0.8mm_comparison_s_index},\ref{fig:CaII_ALMA_0.3mm_comparison_IRT_index},~and~\ref{fig:CaII_ALMA_0.8mm_comparison_IRT_index} (panels c and f) as empty triangles. The data points are near the linear fit and they are very close to the respective VAL~C data point (the green star) in all four cases.

More active stars can exhibit much larger fractional areas of starspots and accordingly larger variations of the S-index, even to an extent where the impact of starspots grows relative to the impact of plage regions \citep{2023ApJ...956L..10S}. A corresponding increase of the continuum brightness temperature for higher sunspot coverage is seen for wavelengths longer than 1.5\,mm. Stellar millimetre observations can therefore provide complementary constraints for the activity level of a star and its variation with a stellar activity cycle.

\section{Conclusions}
\label{Sec:Conclusions}

We investigated synthetic Ca II indices and mm brightness temperatures from a 3D Bifrost model that resembles an enhanced network region surrounded by quiet Sun. Reducing the spatial resolution of these synthetic observables to ALMA resolution decreases the standard deviations of the synthetic observables, leading to increased correlations between them. As the wavelength increases, the beam size also increases, causing small structures to become blurred out and further increasing the correlation between the two diagnostics. However, this decrease in spatial resolution has no significant effect on the mean slopes of the linear fits on the Ca II indices versus mm brightness temperature distributions. This suggests that the instrumental resolution of ALMA is sufficient to capture the similarities between the two diagnostics. The Ca II indices and mm brightness temperatures are closely linked, forming in similar layers in the atmosphere for a wavelength range of 0.3-1.0~mm and providing constraints on the thermal stratification of a star. Previous studies have shown that using multi-wavelength observations alongside mm observations can provide better insights into the (average) thermal structure of stellar atmospheres \citep[][see also references therein]{2023A&A...673A.137P, 2021A&A...655A.113M, 2016SSRv..200....1W}.

Although the Ca II line cores are formed in the chromosphere, the s index, and IRT index are not solely indicative of chromospheric activity. In our model, the s index, for example, reveals mid-to-upper photospheric and lower chromospheric features, such as reverse granulation and internetwork structures. ALMA continuum brightness temperatures are on average too low in these models according to previous studies \citep[see, e.g.,][]{2020ApJ...891L...8M}. The Ca-based activity indices calculated from this model exhibit a notable similarity and exhibit the highest correlation with the ALMA Band~10 data. Therefore, these indicators of stellar activity are heavily influenced by photospheric emission, even though the line cores are formed in the chromosphere. Notably, these indices show clear differences based on activity level, as the values for the QS and EN regions are significantly different.

The superior sensitivity of millimetre continuum observations towards chromospheric structures gives a 3D perspective of the chromosphere in a better manner. Hence, mm brightness temperatures can be used as an alternative and complementary diagnostic to other chromospheric activity indicators with different sources for the opacities.

\begin{acknowledgements}
We thank the referee for very valuable comments and suggestions, which improved considerably the quality of this paper. 
This work is supported by the Research Council of Norway through the EMISSA project (project number 286853), the Centres of Excellence scheme, project number 262622 (``Rosseland Centre for Solar Physics''). Computational resources have been provided by Sigma2 – the National Infrastructure for High-Performance Computing and Data Storage in Norway.
This paper makes use of the following ALMA data: ADS/JAO.ALMA\#2011.0.00020.SV. ALMA is a partnership of ESO (representing its member states), NSF (USA) and NINS (Japan), together with NRC (Canada) and NSC and ASIAA (Taiwan), and KAS (Republic of Korea), in cooperation with the Republic of Chile. The Joint ALMA Observatory is operated by ESO, AUI/NRAO and NAOJ. 
The authors thank \citet{2008A&A...488..653P} and \citet{2016A&A...595A..11L} for their prompt correspondence regarding the Ca II IRT observational data of $\alpha$ Cen A.
The 1D semiempirical \citet{1994ASIC..433..169S} sunspot model was kindly provided by \citet{2014A&A...561A.133L}. 
This research utilised the Python libraries matplotlib \citep{2007CSE.....9...90H}, seaborn \citep{Waskom2021} and the NumPy computational environment \citep{citeulike:9919912}. 
The authors would like to thank Mikolaj Szydlarski, Kjell Andresen, Thore Espedal Moe, and Henrik Eklund for their support.

\end{acknowledgements}

\bibliographystyle{aa}
\bibliography{main}
\end{document}